\documentclass[journal,10pt]{IEEEtran}
\IEEEoverridecommandlockouts
\usepackage{cite}
\usepackage{amsmath,amssymb,amsfonts}
\usepackage{algorithm}
\usepackage{algorithmic}
\usepackage{graphicx}
\usepackage{subcaption}
\usepackage{textcomp}
\def\BibTeX{{\rm B\kern-.05em{\sc i\kern-.025em b}\kern-.08em
		T\kern-.1667em\lower.7ex\hbox{E}\kern-.125emX}}    

\usepackage{amsmath,amssymb,amsfonts,mathrsfs,amsthm,graphicx,epstopdf,bm}
\usepackage{esint,epsfig,soul} %subfigure
\usepackage{cite, color,kotex, comment}
\usepackage[shortlabels]{enumitem}
\usepackage[usenames,dvipsnames,svgnames]{xcolor}
\usepackage{bbm}
\usepackage{multirow}

\newcommand{\problem}[1]{problem (\textsf{P#1})}

\DeclareMathOperator*{\minimize}{minimize}
\DeclareMathOperator*{\maximize}{maximize}

\theoremstyle{definition}

\newcommand{\teq}{\triangleq}

\newcommand\tsum{\textstyle\sum\nolimits}

\newcommand{\gray}[1]{{\color{gray}{#1}}}
\renewcommand{\gray}[1]{}

\newcommand{\E}{\mathbb{E}}
\newcommand{\C}{\mathbb{C}}

\newcommand{\bv}{\mathbf{v}}
\newcommand{\bn}{\mathbf{n}}

\newcommand{\bd}{\mathbf{d}}

\newcommand{\bH}{\mathbf{H}}

\newcommand{\by}{\mathbf{y}}

\newcommand{\sF}{\mathsf{F}}
\newcommand{\G}{\mathcal{G}}

\renewcommand{\P}{\mathscr{P}}

\newcommand{\I}{\mathcal{I}}

\newcommand{\romanum}[1]{\uppercase\expandafter{\romannumeral#1\relax}}

\newcommand{\NN}{\nonumber}

\begin{document}
	
	\title{
%		Wireless Embedded Index Coding with Multi-group Multicast Beamforming via Distributed Multi-Agent Reinforcement Learning 
Joint Design of Embedded Index Coding and  Beamforming for MIMO-based  Distributed Computing via Multi-Agent Reinforcement Learning
	}
	\author{Heekang~Song,~\IEEEmembership{Student~Member,~IEEE},
		and~Wan~Choi,~\IEEEmembership{Fellow,~IEEE}
		% <-this % stops a space
		%\thanks{
		%	This work was supported by Institute of Information \& communications Technology Planning \& evaluation (IITP) grant funded by the Korea government (MSIT) (No. 2018-0-00809, Development on the disruptive technologies for beyond 5G mobile communications employing new resources).
		%}
		\thanks{
			H. Song is with the School of Electrical Engineering, Korea Advanced Institute of Science and Technology (KAIST), Daejeon 34141, Korea  (e-mail: hghsong@kaist.ac.kr). 
		}
		\thanks{	
			W.~Choi is with  the Department of Electrical and Computer Engineering, Seoul National University (SNU), and  the Institute of New Media and Communications, SNU, Seoul 08826, Korea (e-mail:   wanchoi@snu.ac.kr). (\emph{Corresponding author: Wan Choi}).
		}
	}
	\maketitle
	%\vspace{-3}
	
	\begin{abstract}
		In distributed computing systems, reducing the communication load during the data shuffling phase is a critical challenge, as excessive inter-node transmissions are a major performance bottleneck. One promising approach to alleviate this burden is Embedded Index Coding (EIC), which exploits cached data at user nodes to encode transmissions more efficiently. However, most prior work on EIC has focused on minimizing code length in wired, error-free environments—an objective often suboptimal for wireless multiple-input multiple-output (MIMO) systems, where channel conditions and spatial multiplexing gains must be considered. This paper investigates the joint design of EIC and transmit beamforming in MIMO systems to minimize total transmission time, an NP-hard problem. We first present a conventional optimization method that determines the optimal EIC via exhaustive search. To address its prohibitive complexity and adapt to dynamic wireless environments, we propose a novel, low-complexity multi-agent reinforcement learning (MARL) framework. The proposed framework enables decentralized agents to act on local observations while effectively managing the hybrid action space of discrete EIC selection and continuous beamforming design. Simulation results demonstrate that the proposed MARL-based approach achieves near-optimal performance with significantly reduced complexity, underscoring its effectiveness and practicality for real-world wireless systems.    
	\end{abstract}
	
	\begin{IEEEkeywords}
		Distributed computing, Embedded index coding, Spatial multiplexing, Multi-agent reinforcement learning.
	\end{IEEEkeywords}
	%%%%%%%%%%%%%%%%%%%%%%%%%%%%%%%%%%%%%%%%%%%%%%%%%%%%%%%%%%%%%%%%%%%
	% % % % % % % % % % % % % % % % % % % % % % % % % % % % % % % % % %
	%%%%%%%%%%%%%%%%%%%%%%%%%%%%%%%%%%%%%%%%%%%%%%%%%%%%%%%%%%%%%%%%%%%
	\section{Introduction}
	\IEEEPARstart{I}{N} recent years, distributed computing for large-scale data processing has emerged as a dominant computation paradigm, driven by the rapid growth of machine learning and big data applications. Frameworks such as MapReduce \cite{Dean2008} and Spark \cite{Zaharia2010} have gained widespread adoption, where massive computational tasks are divided into two stages: the Map and Reduce phases. In the Map phase, distributed nodes compute functions over locally assigned files and generate intermediate values. To complete their designated functions in the Reduce phase, these nodes must then exchange intermediate values that cannot be computed locally—a process known as the Shuffle phase. Finally, the Reduce phase aggregates the required intermediate values to produce the designated outputs. Notably, in such MapReduce-like systems, the Shuffle phase dominates execution time; for example, \cite{Zhang2013} reports that shuffling accounts for approximately 70\% of the total runtime when performing a SelfJoin on an Amazon EC2 cluster.
	
	%%%%%%%%%%%%%%%%%%%%%%%%%%%%%%%%%%%%%%%%%%%%%%%%%%%%%%%%
 
    To address the communication bottleneck inherent in the Shuffle phase, \textit{Coded Distributed Computing} (CDC)~\cite{Li2018} leverages a fundamental tradeoff between computation and communication. By intentionally increasing redundant computations, CDC creates opportunities to encode data (e.g., via bitwise XOR) such that a single coded transmission can simultaneously satisfy the demands of multiple destinations, thereby reducing the overall communication load. Since its introduction, CDC has been generalized to various operational settings, including straggler-resilient schemes \cite{strLi2016, strYan2019}, wireless networks \cite{wirLi2017, wirLi2019}, topology-aware designs \cite{topWan2020}, and systems with heterogeneous workers \cite{hetkiamari2017, hetwoolsey2021, hetxu2019}.
    
    While prior CDC studies have proposed valuable solutions in specific settings, these approaches are often limited in applicability to their respective environments. Since critical variables such as Map file allocation and Reduce function assignment are strongly coupled with the shuffling design, establishing a structural understanding of data shuffling is essential. In this regard, the \textit{Embedded Index Coding (EIC)} problem, which is known as a NP-hard problem, provides a general formulation of the data shuffling process, capturing its fundamental coding opportunities \cite{Porter2021}. This connection was further established in \cite{Song2022}, which discussed how an EIC solution can be translated into a practical data shuffling design.
		
The EIC problem can be regarded as a device-to-device (D2D) extension of the classical Index Coding (IC) problem~\cite{Bar-Yossef2011}, also known as coded multicasting. In the conventional IC setting, a central sender exploits users' cached side information to encode transmissions, often through XOR operations, so that multiple demands can be simultaneously satisfied. This concept has gained increasing importance with the growing storage capabilities of user devices and the escalating need for efficient multicast strategies to handle the exponential rise in wireless data traffic, projected to reach 120~exabytes by 2026~\cite{Eric2019}. Building on this foundation, the EIC framework extends IC to D2D scenarios in which each node can serve as both a transmitter and a receiver. The key distinction between EIC and the conventional IC problem~\cite{Bar-Yossef2011} lies in the presence of a central server that stores the entire file library. Consequently, in the EIC problem, each user's side information is leveraged not only for decoding but also for encoding messages. For example, recent work such as~\cite{Song2022} provides an optimal-length EIC design for wired communication environments.

    Despite these advances, most IC and EIC studies, such as those by \cite{Bar-Yossef2011, Porter2021, Song2022}, have focused on wired, error-free links, where the optimal code is determined solely by minimizing code length.  However, the growth of wireless distributed computing is driven by the need for multiple devices to collaborate on large tasks in applications like augmented reality (AR), Internet of Things (IoT) networks, and wireless data centers \cite{FLi2019, Yang2019}. In these scenarios, the wireless shuffle phase—the exchange of intermediate results like rendering data in AR or sensor analytics in IoT—is the key bottleneck. This creates a natural opportunity for wireless EIC to improve efficiency by leveraging each node's local data as side information. For instance, \cite{Thomas2015} and \cite{Son2019} show that fading, interference, and channel conditions significantly affect performance of IC problem under single-input single-output broadcast settings. Consequently, this paper is driven by a fundamental question: 
	\begin{itemize}
		\item \emph{Does the pursuit of minimizing the length of index codes remain effective in wireless environment?}
	\end{itemize} 
	In the wireless extension of the EIC problem, termed the \emph{wireless embedded index coding} (WEIC) problem, the emphasis in wireless communications extends beyond merely minimizing the length of the coded scheme to tackling the distinct challenges posed by wireless environments. 
    The efficacy of the minimum length index code is often compromised by inherent characteristics such as channel states, fading, and noise, which do not guarantee superior transmission. Consequently, the priority shifts towards reducing transmission time, a critical performance metric in wireless systems.

    While index coding provides multicast opportunities at the network level, the physical layer offers additional gains through multiplexing techniques such as multicast beamforming. In the WEIC problem, where index-coded messages are transmitted to multiple destinations over wireless channels, this motivates a cross-layer design that jointly exploits network-level multicasting and physical-level multiplexing. %Building on this observation, our focus is on the WEIC problem in MIMO systems. 
    However, extending index coding to multiple-antenna systems is even more challenging due to the NP-hardness of code construction. More recently, \cite{Son2021} demonstrated that spatially multiplexed index coding in MISO broadcast channels can substantially reduce transmission time for wireless IC problem. Motivated by these insights, this paper investigates the joint design of EIC structures and multi-group multicast beamformers in MIMO-based D2D environments, with the goal of minimizing the total transmission time. 
    
    %%%%%%%%%%%%%%%%%%%%%%%%%%%%%%%%%%%%%%%%%%%%%%%%%%%%%%%%%%%%%%%%%%%%%%%%%%%%%%
	\begin{figure}[t]
		\centering
		\begin{subfigure}[b]{0.75\columnwidth}
			\includegraphics[width=\linewidth]{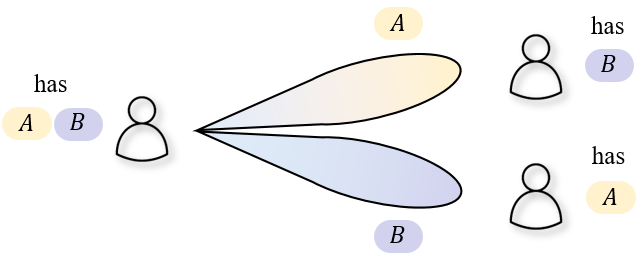}
			\caption{}
			\label{fig:sys1}
		\end{subfigure}
		\hfill 
		\begin{subfigure}[b]{0.75\columnwidth}
			\includegraphics[width=\linewidth]{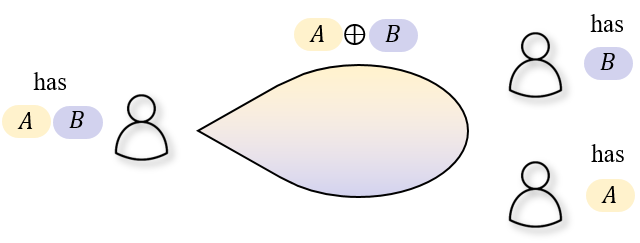}
			\caption{}
			\label{fig:sys2}
		\end{subfigure}
		
		\caption{(a) Spatial multiplexing and (b) Coded multicasting.}
		\label{fig:motiv} 
		\vspace{-10pt}
	\end{figure}
	%%%%%%%%%%%%%%%%%%%%%%%%%%%%%%%%%%%%%%%%%%%%%%%%%%%%%%%%%%%%%%%%%%%%%%%%%%%%%%
	We initiate our discussion with an illustrative example of the WEIC problem within MIMO systems to frame our study's objectives. Referencing Fig. \ref{fig:motiv}, two receivers each request data $A$ and $B$, respectively, while also caching the data requested by the other. The transmitter holds both datasets. In the context of the EIC problem, transmitting a simultaneously encoded message, such as $A \oplus B$ with the XOR operator $\oplus$, proves more beneficial than employing an uncoded unicast. The receivers can subsequently decode their required data by XORing it with the cached data; for example, $(A \oplus B) \oplus A \rightarrow B$ or $(A \oplus B) \oplus B \rightarrow A$.
	
	In the scenario of WEIC problem leveraged by MIMO technology, spatial multiplexing can be leveraged to wirelessly transmit data utilizing beamforming techniques. This introduces two transmission methods: \emph{Spatial multiplexing} and \emph{coded multicasting}. When the channel states of the two receivers are uncorrelated, spatial multiplexing, which involves transmitting spatially multiplexed signals of $A$ and $B$ without encoding, may surpass coded multicasting, exemplified by the index coded signal $A \oplus B$, in reducing transmission time. These findings underscore the need for a joint optimization framework synergizing network-layer coding and physical-layer spatial multiplexing, based on user caches and wireless channel states. The goal is to minimize transmission time in WEIC problem by using a strategic spatial multiplexing gain.

	%%%%%%%%%%%%%%%%%%%%%%%%%%%%%%%%%%%%%%%%%%%%%%%%%%%%%%%%%%%%%%%%%%%
	
	\subsection{Our Contributions and Organizations}    
	The main contributions of this work are summarized as follows.
	\begin{itemize}
		\item We introduce the wireless embedded index coding problem in MIMO networks motivated by distributed computing scenarios. The formulation jointly considers network-layer embedded index coding and physical-layer multi-group multicast beamforming to minimize the total transmission time, explicitly incorporating both coding opportunities and spatial multiplexing gains.
		\item We develop a two-step conventional optimization approach that first designs multi-group multicast beamformers using a conjugate gradient method for a given index code, and then exhaustively searches all feasible EICs to identify the optimal solution. This integrated design approach enables to reduce  transmission time than sequential EIC–beamformer designs. 
		\item To mitigate the prohibitive complexity of exhaustive search and enable efficient operation in dynamic wireless environments, we propose a distributed multi-agent reinforcement learning (MARL) framework. Operating under the centralized training with decentralized execution (CTDE) paradigm, the framework allows decentralized agents to act on local observations and manages the hybrid discrete-continuous action space through a latent action representation using a conditional variational autoencoder (cVAE).
		\item We evaluate the proposed methods through simulations, showing that the MARL-based framework achieves near-optimal performance with substantially reduced computational complexity compared to the exhaustive search baseline. These results demonstrate the effectiveness and practicality of the proposed approach for real-world WEIC scenarios.
	\end{itemize}
	
	The rest of this paper is organized as follows. 
    % Section \ref{sec:PRE} elaborates on the key concepts of EIC and spatial multiplexing, which are foundational to our joint design. 
    Section \ref{sec:SYS} describes the system model for the wireless EIC problem within a MIMO framework and formally presents the optimization problem of jointly designing the EIC and beamforming vectors to minimize transmission time. In Section \ref{sec:CON}, we propose a conventional optimization approach that first designs the beamformer for a given EIC and then searches for the optimal EIC structure. To overcome the complexity and practical challenges of the conventional method, Section \ref{sec:MUL} introduces a novel multi-agent reinforcement learning framework. Section \ref{sec:NUM} numerically evaluates and compares the performance of our proposed design. Finally, Section \ref{sec:CONC} concludes this paper. 
	
	Notations: $\Vert \mathbf{v} \Vert$ denote the vector norm of a vector $\mathbf{v}$. $[1:K]$ denote the sequence of integers from $1$ to $K$.
	
	%%%%%%%%%%%%%%%%%%%%%%%%%%%%%%%%%%%%%%%%%%%%%%%%%%%%%%%%%%%%%%%%%%%
	% % % % % % % % % % % % % % % % % % % % % % % % % % % % % % % % % %

	%%%%%%%%%%%%%%%%%%%%%%%%%%%%%%%%%%%%%%%%%%%%%%%%%%%%%%%%%%%%%%%%%%%
	\section{System Model and Problem Description} \label{sec:SYS}
	%%%%%%%%%%%%%%%%%%%%%%%%%%%%%%%%%%%%%%%%%%%%%%%%%%%%%%%%%%%%%%%%%%%
    In the EIC problem, consider a file library $\mathcal{F}$ consisting of $N$ files. These files are distributed among a set of users denoted as $\mathcal{K}$, where the total number of users is $K$ (with $K \leq N$). Each user $k \in \mathcal{K}$ has a specific file demand $d_k \in \mathcal{F}$, which is not available in their local storage $\mathcal{M}_k \subseteq \mathcal{F}$. Therefore, to fulfill each user's file request, data transmission is required from other users who possess the requested files. This process considers the demands $\mathbf{d} = [d_1, \ldots, d_K]$ of all users and the cached information stored in $\mathcal{M}_1, \ldots, \mathcal{M}_K$. It is assumed that each file in $\mathcal{F}$ has a size of $B$ bits, and the requested files of different users are distinct, meaning that $d_i \neq d_j$ for all $i \neq j$. 
    The objective of the EIC problem is to meet all users' demands with minimal transmission through index coding. We define $\mathcal{E} \triangleq \{m_1, \ldots, m_l\}$ as the set of encoded messages that satisfy all users' demands $\bd$ (i.e., $l \leq K$), where  $l$ is the number of encoded messages, i.e., the length of index codes.   To minimize the length of the index codes, i.e., $|\mathcal{E}|$, it is essential to effectively construct the index code utilizing the (cached) side information.

    In this work, we extend this fundamental problem to a practical wireless context, which we term the WEIC problem. Specifically, we model this context as a half-duplex MIMO communication framework, where each user is equipped with a cache and $N_t$ antennas.
    In the half-duplex MIMO D2D communication, there are $K$ users involved, and in each communication round $t \in [1:K]$, only one user can act as the sender. 
	During round $t$, the sending user (referred to as user $t$) serves the other receiving users in $\mathcal{K} \setminus {t}$ by encoding the demands of the dedicated receiving users based on the cached information $\mathcal{M}_t$. It should be noted that round $t$ can be skipped if there are no dedicated receiving users to be served by user $t$.	Then, the encoded messages served by user $t$ are simultaneously transmitted via spatial multiplexing.	The received signal of receiving user $k \in \mathcal{K} \setminus {t}$ during communication round $t$ is denoted as $\by_{k}[t]  \in \C^{N_t \times 1}$ and can be modeled as:
	\begin{align}
		\by_{k}[t]=\bH_{k}^{\dagger}[t] \mathbf{x}[t] + \bn_{k}[t],
	\end{align}
	where $\bH_{k}[t]\in\C^{N_t \times N_t}$ is user $k$'s channel matrix at communication round $t$ and $(\cdot)^\dagger$ denotes the conjugate transpose of a vector.
	We consider a block fading channel model so that the channel coefficient $\bH_{k}[t]$ is a constant during a single transmission period but varies across different transmissions. It is also assumed that the transmitter exactly knows   users' channels, i.e., $\{\bH_{1}[t], \ldots, \bH_{K}[t]\}$. Also, $\mathbf{x}[t] \in \C^{N_t \times 1}$ and $\bn_{k}[t]\in \C^{N_t \times 1}$ are a transmit signal sent by user $t$ and a complex Gaussian noise with zero mean and unit variance, respectively. %, such that $\E[\Vert\mathbf{x} [t]\Vert^2] \le P$, where $P$ is the total transmit power budget for each user.} 
	The transmission is comprised of two components: 1)  EIC and 2) beamforming vectors.
	%
	%Define $\mathcal{E} \triangleq \{m_1, \ldots, m_l\}$ as the set of encoded messages that satisfy all the demands $\bd$ of users (i.e., $l \leq K$). 
	It should be noted that the use of spatial multiplexing allows the sending user to transmit multiple messages simultaneously, significantly reducing the number of required transmission rounds, potentially to fewer than both $l$ and $K$.
	
	Now, we can represent an EIC with $K$ disjoint groups, where each group corresponds to the subset of encoded messages to be transmitted  by each sending user during each communication round.
	Let $\G_t \subset \mathcal{E}, \forall t \in [1:K]$ be the set of encoded messages that are sent simultaneously by  user $t$. Then, $\G_1\cup\ldots\cup\G_K = \mathcal{E}$. As aforesaid, it is possible that the number of required transmission rounds is less than $K$, resulting from some groups being empty sets.
	That is, the transmit signal $\mathbf{x}[t]\in\C^{N_t \times 1}$ at communication round $t$ is constructed as follows:
	\begin{align}
		\mathbf{x}[t]\teq \sum_{i\in \I_t} \bv_i m_i \label{eqn:tx_signal},
	\end{align}
	where $\I_t$ is the set of indices of the encoded messages in $\G_t$, $\mathbb{E}[m_i] =0$ and $\E[\vert m_i\vert^2]=1, \forall i \in [1:l]$. $\bv_i \in \C^{N_t \times 1}$ is a beamforming vector for the encoded message $m_i$ such that $\sum_{i\in \I_t} \Vert \mathbf{v}_i \Vert^2  \leq P$
	where $P$ is the total transmit power budget for each user.
	%Denote $\mathcal{D}_t \subset \mathcal{K}$ as the set of users served at communication round $t$. Since $\mathcal{D}_i$ and $\mathcal{D}_j$ are disjoint for all $i \neq j$, the encoded messages $m_1, \ldots, m_l$ are independent of each other,
	For physical-layer transmission, each file is represented as a sequence of bits. We then independently encode (XORed) bit streams of the requested files into encoded message and the demand of each user is associated with no more than one encoded messages in $m_1, \ldots, m_l$, 
	%i.e., 
	thus
	$
	\E[\Vert\mathbf{x}[t]\Vert^2] \le P, \forall t \in [1:K].
	$
	Note that, if some users are served together, there exists an empty group, and we do not allocate power to the corresponding beam, i.e., $\Vert \bv_i \Vert^2=0$ whenever $\G_i=\emptyset$, implying zero transmission time.

	%%%%%%%%%%%%%%%%%%%%%%%%%%%%%%%%%%%%%%%%%%%%%%%%%%%%%%%%%%%%%%%%%%%
	% % % % % % % % % % % % % % % % % % % % % % % % % % % % % % % % % %
	%%%%%%%%%%%%%%%%%%%%%%%%%%%%%%%%%%%%%%%%%%%%%%%%%%%%%%%%%%%%%%%%%%%
	%\subsection{Problem description} \label{sub:prob_description}
	For a given EIC $\G_\mathcal{K} (\triangleq \G_1, \ldots, \G_K)$ dedicated to the users in $\mathcal{K}$, at each communication round $t$, the transmitter $t$ (i.e., user $t$) allocates transmit power $P$ to $|\G_t|$ different streams (or $|\G_t|$ beamforming vectors) to minimize the transmission time. Define $\mathcal{D}_t \subset \mathcal{K}$ as the set of users served at communication round $t$.
	%
	% Note that the total transmission time is related with the quality of service, which is important in wireless contents delivery networks such as video streaming. 
	Then, the received signal of receiving user $k \in\mathcal{D}_t$ at communication round $t$ becomes
	\begin{align}
		\by_{k}[t]
		& = \bH_{k}^\dagger[t]\bv_{\mathcal{C}(k)}m_{\mathcal{C}(k)}+ \bH_{k}^\dagger[t] \bigg(\sum_{i \in \I_t\backslash \{\mathcal{C}(k)\}}\bv_im_i\bigg)+\bn_{k}[t].
	\end{align}
	where $\mathcal{C}(\cdot)$ returns the index of the encoded message that is required by the receiving user. % group index indicator that takes a user index and returns the group index belonging the user, i.e., $\mathcal{C}(k) = i ~~\longleftrightarrow~~ k\in\G_i.$}

	 The signal-to-interference-plus-noise-ratio (SINR) of receiving user $k$ becomes the function of beamforming vectors $\bv_{\I_t}(\triangleq \bv_i, \forall i \in \I_t)$ given by
	\begin{align}
		\mathsf{SINR}_{tk}(\bv_{\I_t}) \teq \dfrac{\Vert\bH_{k}^\dagger[t]\bv_{\mathcal{C}(k)}\Vert^2}{\sum_{i \in\I_t\backslash \{\mathcal{C}(k)\}}\Vert\bH_{k}^\dagger[t]\bv_{i}\Vert^2+1}, \label{eqn:SINR}
	\end{align}
	and   receiving user $k$'s achievable rate is $R_{tk} \teq W  \log\left(1+\mathsf{SINR}_{tk}(\bv_{\I_t})\right)$, where $W$ is total system bandwidth.
	Thus, the  transmission time required by  transmitter $t$ (i.e., sending user $t$) becomes $ B/R_{tk}$.

	Since  transmitter $t$ serves all encoded messages in $\G_t$ by spatial multiplexing at communication round $t$, the total transmission time required to serve all the users becomes
	\begin{align}
		T \teq \sum_{t = 1}^K T_t
		=\sum_{t = 1}^K \max_{ k \in \mathcal{D}_t}
		\left[\dfrac{B/W}{\log(1+\mathsf{SINR}_{tk}(\bv_{\I_t}))}\right].
	\end{align}
	%
	%\blue{Without loss of generality, to ease of explanation, we assume that the system bandwidth is equal to one, i.e., $W=1$.}
	Notice that the total transmission time is affected by the constructed code design $\G_t$ and their beamformer $\bv_{\I_t}$ for all communication round $t \in [1:K]$.
	
	Thus, we formulate the problem that finds the optimal EIC and the optimal beamforming vectors to minimize the total transmission time under the power constraint as follows: 
	\begin{align}
		(\textsf{P1})~
		\!\!\minimize_{(\G_1, \ldots, \G_K), \atop (\bv_{\I_1}, \ldots, \bv_{\I_K})} ~& T \\%\NNL %\label{eqn:objective}\\
		\textrm{subject to} \!~~
		& \sum_{i\in \I_t} \Vert \mathbf{v}_i \Vert^2  \leq P, \textrm{~~for all~~} t\in[1:K], \label{eqn:constraint1}\\
		& \G_1\cup\ldots\cup\G_K= \mathcal{E}, \label{eqn:constraint2}\\
		& \G_i\cap\G_j= \emptyset \textrm{~~for all~~} i\ne j, \label{eqn:constraint3}\\
		& \mathcal{E} \subseteq \P,
		%\M_k\!\supset\!\{ \R_n ~\!\vert\!~ n\!\in\!\G_{\I(k)} \!\!\setminus\! \{k\} \} \!\!\textrm{~~\!for \!~}\! k\!\in\![1:K],
		\label{eqn:constraint4}
	\end{align}
	where $\P$ is the all possible set of encoded messages for a given $\bd$ and $\mathcal{M}$. 
	% finding the decodable coding clique set $\mathscr{P}|_\mathbf{W}$ for a side information graph, which is called \textit{clique problem}, is known to be NP-hard \cite{ref:karp,Song2022}.
	
	%%%%%%%%%%%%%%%%%%%%%%%%%%%%%%%%%%%%%%%%%%%%%%%%%%%%%%%%%%%%%%%%%%%
	% % % % % % % % % % % % % % % % % % % % % % % % % % % % % % % % % %
	%%%%%%%%%%%%%%%%%%%%%%%%%%%%%%%%%%%%%%%%%%%%%%%%%%%%%%%%%%%%%%%%%%%
	\section{Conventional optimization approach} \label{sec:CON}
	Since the \problem{1} cannot be directly solved, we solve $(\textsf{P1})$ iteratively. First, we design the  beamforming vectors for an arbitrarily given EIC and then exhaustively search the optimal EIC with the designed beamforming vectors.  
	For an arbitrarily given $\G_1,\cdots,\G_K$, the beamformers $\bv_{\I_t}$ at communication round $t$, minimizing the transmission time $T_t$, are independent of the other communication rounds $[1:K] \setminus \{t\}$, due to half-duplex transmission. Moreover, the transmission time at each communication round is a monotonic decreasing function of $\mathsf{SINR}$. As a result, given $\G_1,\cdots,\G_K$, the \problem{1} is equivalently resolved by optimizing the subproblems $(\textsf{SP1})$ for every communication round $t\in[1:K]$,  
	\begin{align}
		(\textsf{SP1})~
		\maximize_{\bv_{\I_t}} \quad& \min_{k\in \mathcal{D}_t} \mathsf{SINR}_{tk}(\bv_{\I_t}) \label{prob:SINR2}\\
		\textrm{subject to}
		\quad&
		%\sum_{i\in \I_t} \Vert \mathbf{v}_i \Vert^2  \leq P
		\eqref{eqn:constraint1}
		%\eqref{eqn:constraint5}
		.\NN
	\end{align}
	Since the beamforming vectors $\bv_{\I_t}$ is for the encoded messages in the group $\G_t$, the problem $(\textsf{SP1})$ can be regarded as a general \emph{max-min fair multi-group multicast beamforming problem}, which is known as NP-hard \cite{Karipidis2008}.
	Therefore, we propose a conjugate gradient based multi-group multicast beamforming vectors \cite{Zhou2017, Son2021}.

	\subsection{Conjugate Gradient based Multi-group Multicast Beamformer  \label{Sec:conj_beamformer}}
	Since the procedure is the same for all communication round $[1:K]$, here we focus on the subproblem $(\textsf{SP1})$ at  round $t$. The minimum SINR in \eqref{prob:SINR2} is represented as
	\begin{align}
		\min_{k \in \mathcal{D}_t} \left[ 
		\Vert \bH_{k}^{\dagger}[t]  \bv_{\mathcal{C}(k)}\Vert^2 \middle/\left(\sum_{i \in\I_t\backslash \{\mathcal{C}(k)\}}\Vert\bH_k^{{\dagger}}[t] \bv_{i}\Vert^2+1 \right)
		\right]. %\label{eqn:det_SINR}
	\end{align}
	
	Thus, to apply the gradient descent algorithm, $(\textsf{SP1})$ is transformed into an equivalent parametric programming problem with an auxiliary parameter $\eta$ \cite{Dinkelbach1967}. Thus, we transform the objective into $\sF(\bv_{\I_t},\!\eta) \!\teq\! \min_{k} \! \sF_k (\bv_{\I_t},\!\eta)$,	where 
	\begin{equation}
		\sF_k(\bv_{\I_t},\eta) \teq \Vert\bH_k^{\dagger}[t] \bv_{\mathcal{C}(k)}\Vert^2 -\eta \cdot \left(\sum_{i \in\I_t\backslash \{\mathcal{C}(k)\}}\Vert\bH_k^{{\dagger}}[t]\bv_{i}\Vert^2\!+\!1 \!\right).
	\end{equation}
	Due to a non-smoothness of $\sF_k(\bv_{\I_t},\!\eta)$, using the log-exp smoothing function, we change the objective function as
	\begin{align}
		\mathit{F}(\bv_{\I_t},\eta,\mu) \teq \mu\log\bigg(\tsum_{k\in\mathcal{D}_t}\exp (- \sF_{k}(\bv_{\I_t},\eta)/\mu) \bigg)
	\end{align} and  rewrite the problem $(\textsf{SP1})$ as
	\begin{align}
		(\textsf{SP2})~
		\maximize_{\bv_{\I_t}} \quad& \mathit{F}(\bv_{\I_t},\eta,\mu) \\
		\textrm{subject to}
		\quad&
		\eqref{eqn:constraint1}
		%\eqref{eqn:constraint5}
		.\NN
	\end{align}
	Here, $\mu$ is a smoothing parameter.
	We can solve the problem $(\textsf{SP2})$ by using the Dinkelbach-type Riemannian conjugate gradient (DT-RCG) algorithm \cite{Zhou2017}, which is a gradient descent algorithm for manifold optimization. % We define $\bv_{1:K}^{\star}$ as the corresponding conjugate gradient beamformer.
	
	\subsection{Optimal Embedded Index Code Search}
	To obtain the optimal solution for the problem $(\textsf{P1})$, the straightforward approach is to exhaustively search all the EICs by solving the subproblem $(\textsf{SP2})$ for all communication round $t \in [1:K]$. Since the encodable/decodable EIC set cannot be structured, the first step is to identify all feasible EICs among the potential options for a given cached information $\mathcal{M}$. Subsequently, we iteratively compare the total transmission time $T$ for the feasible EICs to determine the best solution. The detailed algorithm is illustrated in Algorithm \ref{alg:es}.
	
	\begin{algorithm}[t]
		%\setstretch{0.7}
		\caption{Overall Conventional Optimization Approach} %The Optimal Embedded Index Code Search}
		\begin{algorithmic}
			\STATE Construct the possible encoded message set $\mathscr{P}$
			\STATE Solve subproblem $(\textsf{SP2})$ and compute transmission time $T$ for all EIC $\mathcal{E} \subseteq \mathscr{P}$
			\STATE Find the minimum time $T^*$ and its EIC $\mathcal{E}^*$
		\end{algorithmic}
		\label{alg:es}
	\end{algorithm}

	Due to the absence of closed-form solutions for expressing the beamforming vector as a multicast beamformer, each EIC requires exhaustive searching. However, the favorable aspect is that the number of feasible EICs is finite in the case of clique cover index coding\footnote{Clique cover index coding is a graph-based method where messages and side information are represented as a graph. The minimum number of transmissions corresponds to the minimum number of cliques required to cover all vertices in that graph \cite{Song2022}.}. Consequently, a finite search over the feasible EICs can lead to the solution for \problem{1} given cached information $\mathcal{M}$.

	\subsection{Complexity Analysis}
	
	We shall briefly explore the computational complexity of our proposed schemes using big-O notation, considering the worst-case scenario. With the EIC problem transformed into multiple IC problems and each node serving as a central server during their communication rounds, the maximum count of decodable index codes in each communication round reaches up to $2^K$. Treating up to $K$ communication rounds as independent, the total number of possible EICs is thus bounded by $O((2^K)^K)=O(2^{K^2})$, since each round may admit up to $2^K$ candidate index codes and these choices compound multiplicatively across the $K$ rounds. For DT-RCG approach, it is essential to compute the conjugate gradient in every iteration, a task with a complexity of $O(K^3N_t)$ \cite{Zhou2017}, considering a cap of $N_\text{iter}$ iterations. In summary, the cumulative computational complexity of our proposed methodology becomes $O(2^{K^2} \cdot N_\text{iter} K^3N_t)$.
	
	%%%%%%%%%%%%%%%%%%%%%%%%%%%%%%%%%%%%%%%%%%%%%%%%%%%%%%%%%%%%%%%%%%%
	% % % % % % % % % % % % % % % % % % % % % % % % % % % % % % % % % %
	%%%%%%%%%%%%%%%%%%%%%%%%%%%%%%%%%%%%%%%%%%%%%%%%%%%%%%%%%%%%%%%%%%%
	% \section{Distributed Reinforcement learning approach}
	\section{Distributed Multi-agent Reinforcement learning approach} \label{sec:MUL}
	The conventional two-step optimization approach proposed for the WEIC problem encounters significant difficulties due to the exponential growth of the search space.  Furthermore, in conventional centralized optimization approaches, acquiring global CSI (i.e., the channel states for all D2D pairs) at a single controller is practically challenging because of substantial signaling overhead, and this CSI often varies dynamically over time. Such variations can increase the transmission time of optimally designed EIC and multi-group multicast beamforming, especially in dynamic wireless environments where CSI may become outdated before the time-consuming centralized optimization process is completed. As a result, relying on instantaneous, globally shared CSI is often practically suboptimal. 
	To overcome these limitations, we propose a distributed multi-agent reinforcement learning (MARL) framework. While both approaches require CSI to construct beamforming vectors, our key contribution lies in removing the need for a centralized entity. Our framework enables each agent to make fast, decentralized decisions based only on its locally observed CSI (i.e., the channels from itself to other users). This not only mitigates the exponential search complexity but also enhances robustness against stale CSI by avoiding the bottleneck of global information aggregation, ensuring near-optimal performance in rapidly changing environments.   
	 %, which enables each agent to act based only on local observations of CSI. This not only mitigates the exponential search complexity but also enhances robustness against stale or infrequent CSI feedback, ensuring near-optimal performance even in rapidly changing environments. \red{\bf XXX I cannot understand the CSI argument. Both the conventional scheme and the MARL algorithm requires CSI to construct bemaforming vectors XXX}
	
	To facilitate the application of distributed MARL algorithms, we recast the original problem $(\textsf{P1})$ within the framework of a Partially Observable Markov Decision Process (POMDP). At communication round $t \in \mathcal{K}$, agent $t$ observes the pending requests, together with the locally known CSI $\bH[t]\triangleq{\bH_{1}[t], \ldots, \bH_{K}[t]}$ between itself and the other agents. We assume channel reciprocity, such that uplink and downlink channels share the same propagation paths. The agent also observes the cached information $\mathcal{M}$. Then, all the available information  is formally represented as
	\begin{equation}
		\mathcal{O}_t = \{\mathbf{e}_t, \bH[t], \mathcal{M} \},
	\end{equation}
	where $\mathbf{e}_t \in [0, 1]^K$  represents the status of remaining requests after communication round $t-1$, depicted as a binary indicator vector, and the cached information $\mathcal{M}$ can be transformed into the adjacency matrix of side information graph \cite{Song2022}. Each element in the indicator vector $\mathbf{e}_t$ corresponds to whether a user’s request has been satisfied: ‘1’ indicates a fulfilled request, while ‘0’ denotes an unfulfilled one. Based on this observation, the agent $t$ determines the action as
	\begin{equation}
		\mathcal{A}_t = \{\G_t, \bv_{\I_t}\}.
	\end{equation}
	After all agents execute their actions $\{ \mathcal{A}_1, ..., \mathcal{A}_K \}$, the environment provides a team-wide reward defined as 
	\begin{equation} \label{eq:reward}
		\mathcal{R} = 1/T,
	\end{equation} 
	based on the total transmission time $T$. Note that minimizing the overall transmission time is equivalent to maximizing the sum of rewards over all agents.  
	
	However, two critical challenges arise when directly applying DRL algorithms in this setup.
	\begin{enumerate}
		\item The \textit{hybrid-action space problem} arises because the EIC action $\G_t$ lies in a discrete space while the beamforming action $\bv_{\I_t}$ is continuous. Conventional DRL methods typically handle either discrete or continuous actions separately, making them ill-suited for our setting.
		\item The \textit{credit assignment problem} arises because individual contributions to the time‑based reward are hard to isolate. When multiple users can serve a request, letting the best channel user transmit (and others refrain) reduces overall time, yet the non‑transmitting agent receives no reward, revealing a misalignment between team benefit and individual incentive.
	\end{enumerate}
	
	Our framework resolves the above challenges through two key components. First, to overcome the credit assignment problem, where the contribution of each agent’s action to the global transmission time is ambiguous. we adopt the centralized training with decentralized execution paradigm, enhanced with an attention-based centralized critic that delivers individualized learning signals to each agent. Second, to address the hybrid-action space problem, arising from the coexistence of discrete EIC actions $\G_t$ and continuous beamforming actions $\bv_{\I_t}$, we propose a latent policy learning scheme inspired by HyAR \cite{Li2022}. Specifically, discrete actions are represented by a learnable embedding table, in which each row corresponds to a continuous vector representation of a discrete action index, while continuous actions are generated through a cVAE decoder. Collectively, these components render our MARL framework effective in dynamic wireless environments.
	
	\subsection{Overall Framework of the Proposed MARL Algorithm}
	Our proposed MARL algorithm is founded on an actor-critic architecture, tailored to address the unique challenges of the WEIC problem. The framework consists of multiple decentralized actors, one for each agent, and a single centralized critic. Each actor, observing only its local state $\mathcal{O}_t$, learns a policy in a latent space to address the hybrid-action problem. Under the CTDE paradigm, agents act sequentially, yet the centralized critic jointly evaluates all actions for cooperative team optimization, using an attention mechanism to resolve credit assignment with individualized feedback. The entire framework is integrated with the Twin Delayed Deep Deterministic Policy Gradient (TD3) algorithm for stable and efficient policy learning \cite{Fujimoto2018}.
	
	\subsubsection{Latent Policy-based Actors for Hybrid Actions}
	To tackle the hybrid-action space challenge, where the action $\mathcal{A}_t = \{\G_t, \bv_{\I_t}\}$ comprises both discrete and continuous components, we leverage a latent action framework inspired by HyAR \cite{Li2022}. Each actor network $\pi_t(\cdot|\mathcal{O}_t)$ outputs a continuous latent action $z_t = [z_t^d, z_t^c]\in\mathbb{R}^{|z_t^d|+|z_t^c|}$, which is subsequently decoded into the original hybrid action space.
	
	The discrete component of the latent action, $z_t^d$, determines the EIC action. This is achieved by identifying the index of the closest row in a learnable embedding table $E_{\zeta} \in \mathbb{R}^{D \times |z_t^d|}$:
	\begin{equation}
		%\G_t 
		g_t = \underset{i \in \{1, \dots, D\}}{\arg\min} \| z_t^d - E_{\zeta}(i) \|_2^2,
	\end{equation}
	where $E_{\zeta}(i)$ denotes the $i$-th row of the embedding table, and $D$ denotes the total number of possible actions to serve the maximum $N_t$ users.
For example, in communication round $t=1$, the agent can either remain silent, send an index-coded message to users $\{2,3\}$, or simultaneously transmit separate messages to users $\{2,4\}$ through spatial multiplexing. 	In general, $D$ is obtained by selecting a subset of $N_t$ users and partitioning it into simultaneously transmitted blocks. Since the action space includes all subsets of the $N_t$ receivers (including the empty subset representing a skip transmission), its size is $D=\sum_{s=0}^{N_t}\binom{N_t}{s} \mathscr{B}_s = \mathscr{B}_{N_t+1}$, where $\mathscr{B}_s$ denotes the $s$-th Bell number.\footnote{User clustering can mitigate scalability issues by grouping users either by side-information overlap (for coding gain) or spatial proximity (for channel quality), thereby reducing the effective action space $D$.} Note that when computing $g_t$, we rely on the local observation, specifically, the request status $\mathbf{e}_t$, by masking fulfilled requests and selecting the closest action for the remaining ones. Based on the selected index $g_t$, we map $g_t$ to the corresponding set of encoded messages $\G_t$.
	
	The cVAE consists of an encoder and a decoder.  
	The encoder $q_{\phi}(z_t^c \mid \bv_{\I_t}, \mathcal{O}_t, E_{\zeta}(g_t))$ maps the continuous parameter $\bv_{\I_t}$ together with the local observation $\mathcal{O}_t$ and the embedding of the chosen discrete action $E_{\zeta}(g_t)$ into a Gaussian latent variable $z_t^c$, where the mean and variance are outputs of the encoder network. 
	Then, the decoder employs this latent variable to realize the latent policy by reconstructing the continuous beamformer, conditioned again on $\mathcal{O}_t$ and $E_{\zeta}(g_t)$:
	\begin{equation}
		\tilde{\bv}_{\I_t} = p_{\psi}(z_t^c, \mathcal{O}_t, E_{\zeta}(g_t)).
	\end{equation}
	The reconstructed vector $\tilde{\bv}_{\I_t}$ is then used  for transmission. During warm-up training, the encoder uses a reference beamformer for $\bv_{\I_t}$ obtained from a low-complexity solver (e.g., SDR) as a supervised target. Together with the observation and action embedding, this reference guides the shaping of the latent distribution. The reference is used only during warm-up to ensure accurate reconstruction and stabilize representation learning. In the execution phase, the encoder is omitted; instead, the actor generates the latent input, and the decoder produces the transmitted beamformer.

	Conditioning on the embedding of discrete action  is crucial. It allows the cVAE to explicitly model the dependency of the continuous parameters on the discrete choice. Furthermore, as the embedding table $E_{\zeta}$ is trained, it learns to place semantically similar discrete actions closer in the latent space, promoting smoother and more efficient policy learning. This entire mechanism transforms the complex hybrid-action problem into a tractable continuous-space policy learning problem.
	
	\subsubsection{Attention-based Critic for Credit Assignment}
	Our primary objective is to minimize the team-wide total transmission time. To fairly evaluate each actor's policy towards this shared goal and solve the credit assignment problem, we design a centralized critic that incorporates an attention mechanism. Instead of providing a single global value, the critic computes an individualized Q-value, $Q_t$, for each agent $t$, reflecting its unique contribution.
	
	The centralized critic first embeds each agent's local observation and latent action as $\varepsilon_t = f_{\varepsilon}(\mathcal{O}_t, z_t)$. From these embeddings, query, key, and value representations are generated via learnable linear projection matrices  $M_\mathbf{q}, M_\mathbf{k}$, and $M_\mathbf{w}$, respectively:
	$\mathbf{q}_t = M_\mathbf{q} \varepsilon_t, \mathbf{k}_t = M_\mathbf{k} \varepsilon_t, \mathbf{w}_t = M_\mathbf{w} \varepsilon_t.$
	The attention weights for agent $t$ are then computed as
	\begin{equation}
		\alpha_t = \mathrm{softmax}\!\left(\frac{\mathbf{q}_t \mathbf{k}^\top}{\sqrt{l_\mathbf{k}}}\right),
	\end{equation}
	where $\mathbf{k}=[\mathbf{k}_1,\dots,\mathbf{k}_K]^\top$ is the key matrix and $l_\mathbf{k}$ is the key dimension. 
	The context vector for agent $t$ is given by $\alpha_t \mathbf{w}$, where $\mathbf{w}=[\mathbf{w}_1,\dots,\mathbf{w}_K]^\top$ is the value matrix. 
	Finally, the individualized Q-value for agent $t$ is then a function of its own embedding and this context vector, $Q_t = f_Q(\varepsilon_t, \alpha_t \mathbf{w})$. Here, $f_\varepsilon$ and $f_Q$ denote the learnable functions.

	The critic network is trained to minimize the value estimation error based on the sum of these individual Q-values, i.e., $Q =\sum_t Q_t$. Subsequently, each actor $\pi_t$ is updated using the policy gradient derived specifically from its own Q-value, $Q_t$. This ensures that each agent receives a precise and relevant learning signal, enabling the framework to learn effective cooperative strategies where individual contributions, such as refraining from transmission, are appropriately valued.
	
	%%%%%%%%%%%%%%%%%%%%%%%%%%%%%%%%%%%%%%%%%%%%%%%%%%%%%%%%%%%%%%%%%%%%%%%%%%%%%
	\begin{figure*}[t]
		\centering
		{\includegraphics[width=0.9\textwidth]{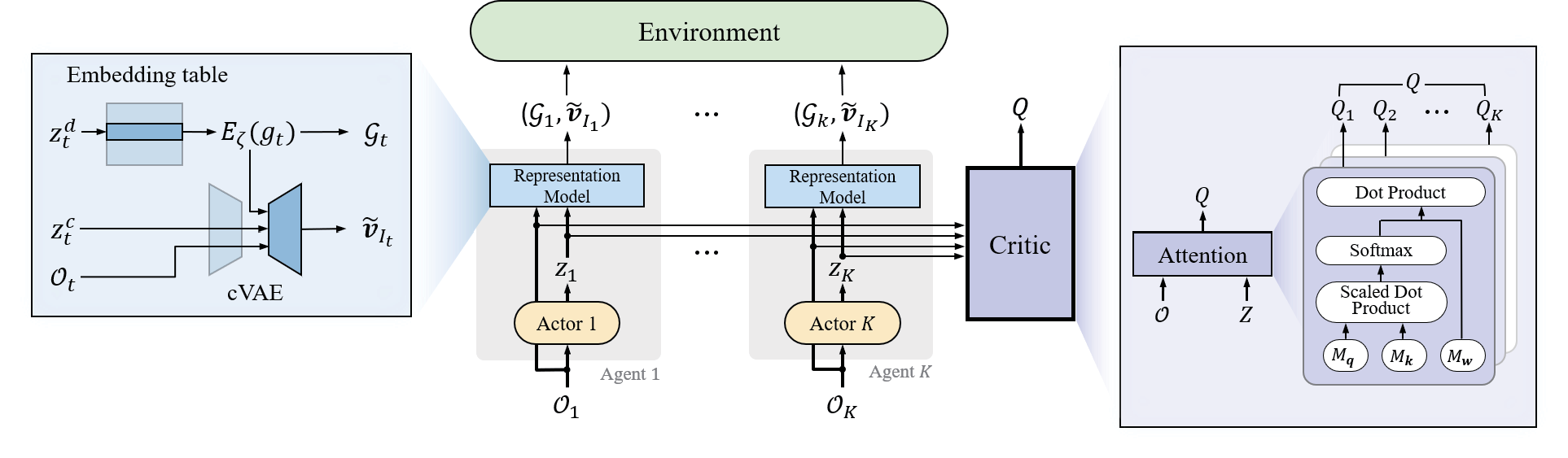}} 
		\caption{\!\!\! {The proposed distributed multi-agent reinforcement learning structure.  }} 
		\label{fig:framework}
	\end{figure*}
	%%%%%%%%%%%%%%%%%%%%%%%%%%%%%%%%%%%%%%%%%%%%%%%%%%%%%%%%%%%%%%%%%%%%%%%%%%%%%%
	The overall structure is depicted in Fig. \ref{fig:framework}.
	
	\subsection{Training Objectives and Network Updates}
	The proposed framework is trained by concurrently optimizing three main components: the hybrid action representation model, the centralized critic networks, and the decentralized actor networks.
	
	\subsubsection{Representation Model Training}
    The representation model, consisting of the embedding table $E_{\zeta}$ and the cVAE parameterized by $\phi$ (encoder) and $\psi$ (decoder), is trained to create a compact and semantically smooth latent space through a two-stage process. Initially, during the warm-up training stage, the cVAE is trained using a supervised target (i.e., reference beamformers) to learn a robust initial mapping. Subsequently, during the main MARL training stage, the model is periodically updated in an unsupervised manner.
	First, the VAE loss, $\mathcal{L}_{\text{vae}}$, ensures that the latent space can accurately reconstruct the original continuous actions. 
	It consists of the L2 reconstruction error and a Kullback-Leibler (KL) divergence term that regularizes the latent space \cite{Li2022}:
	\begin{equation}
		\mathcal{L}_{\text{vae}}(\phi,\psi,\zeta)=\mathbb{E} \left[ \| \bv_{\I_t} - \tilde{\bv}_{\I_t} \|_{2}^{2} + D_\text{KL}(q_{\phi}(\cdot|\cdot)||\mathcal{N}(0,I)) \right]
	\end{equation}
	where $\tilde{\bv}_{\I_t}$ is the reconstructed beamforming vector. 
	
	In addition, to enforce a more nuanced semantic structure, we introduce a specialized semantic similarity loss designed to structure the latent space. By ensuring that semantically similar actions are mapped to nearby vectors in the latent space, we achieve two key benefits. First, it facilitates smoother and more efficient learning for the policy and value functions, as they can generalize knowledge across actions that are likely to yield similar outcomes. Second, it enhances the representation's smoothness, enabling the cVAE decoder to generate consistent continuous actions (i.e., beamformers) for similar discrete action inputs. This loss is structured based on two distinct criteria of semantic similarity:
	\begin{itemize}
		\item \textbf{Destination similarity}: We use a destination-wise InfoNCE loss \cite{Oord2018} to pull the embeddings of discrete actions ($E_{\zeta}(i)$) that serve the same set of destination users closer together, while pushing them apart from others. 
		The rationale is that actions targeting the same users will interact with similar channel conditions and data demands, likely resulting in similar transmission times. Structuring the latent space this way helps the value function to learn this correlation effectively.
		Given two embeddings $E_{\zeta}(i)$ and $E_{\zeta}(j)$, they form a positive pair if their corresponding actions target the identical set of users. The total destination similarity loss is defined as the sum of the InfoNCE losses for all embeddings:
		\begin{equation}
			\mathcal{L}_{\text{ds}} = -\frac{1}{|\mathcal{B}|}\sum_{i \in \mathcal{B}}  \log \frac{\sum_{j \in \text{pos}(i)} \exp(\text{sim}(E_{\zeta}(i), E_{\zeta}(j))/\tau)}{\sum_{k \neq i} \exp(\text{sim}(E_{\zeta}(i), E_{\zeta}(k))/\tau)},
		\end{equation}
		where $\text{pos}(i)$ is the set of indices forming positive pairs with $E_{\zeta}(i)$ for the sample $i$, $\text{sim}(\cdot)$ is the cosine similarity, and $\tau$ is a temperature hyperparameter.
 
        \item \textbf{Composite channel similarity}: To enable the model learn the relationship between channel characteristics and optimal beamformers, we organize the continuous latent space based on composite channel similarity. The composite channel reflects a destination set's spatial signature. The key idea is that users with similar composite channels should be served by structurally similar beamformers. We enforce this by structuring the latent space so that continuous representations ($\mathbf{z}^c_t$) of actions targeting spatially similar groups are clustered together.
        %To enable the model to learn the intrinsic relationship between channel characteristics and optimal beamforming structures, we organize the continuous latent space according to the similarity of composite channels. The composite channel for a given destination set captures its overall spatial signature. The key idea is that users with highly similar composite channels should be served by structurally similar beamformers. To enforce this principle, we structure the latent space such that the continuous representations ($\mathbf{z}^c_t$) of actions targeting spatially similar groups are positioned closer together.
		
		For each sample $j$ in a batch $\mathcal{B}$, we compute its composite channel matrix $\mathbf{S}j$, which represents the averaged spatial characteristics of its destination user set $\mathcal{D}j$:
		\begin{equation}
			\mathbf{S}_j = \frac{1}{|\mathcal{D}_j|} \sum_{k \in \mathcal{D}_j} \frac{(\mathbf{H}_{k}[j])^\dagger \mathbf{H}_{k}[j]}{\text{tr}((\mathbf{H}_{k}[j])^\dagger \mathbf{H}_{k}[j])}.
		\end{equation}
		We then measure the similarity between two composite channels, $\mathbf{S}_i$ and $\mathbf{S}_j$, using the cosine similarity $c_{ij}$. This metric quantifies spatial alignment: a value near 1 indicates high alignment, while one near 0 implies near orthogonality. The fundamental principle is that groups with similar composite channels should require similar beamforming structures.

    Based on physical-layer similarity $c_{ij}$, we define positive pairs (if $c_{ij} \geq t_{\text{pos}}$) and negative pairs (if $c_{ij} \leq t_{\text{neg}}$) relative to an anchor sample $i$. A weighted InfoNCE loss is then applied to attract the latent vectors of positive pairs and repel those of negative pairs. Let $\hat{\mathbf{z}}_i$ denote the continuous latent vector of the $i$-th batch sample. The loss is defined as:
		\begin{equation}
			\mathcal{L}_{\text{ccs}} = -\frac{1}{|\mathcal{B}|}\sum_{i\in\mathcal{B}} \log \frac{\sum_{j\in \text{pos}(i)}\exp(\text{sim}(\hat{\mathbf{z}}_i, \hat{\mathbf{z}}_j)/\tau)}{\sum_{k \neq i} w_{ik} \exp(\text{sim}(\hat{\mathbf{z}}_i, \hat{\mathbf{z}}_k)/\tau)},
		\end{equation}
		where $\text{pos}(i) = \{j \neq i : c_{ij} \geq t_{\text{pos}}\}$ is the set of positive pairs for anchor $i$, $\text{sim}(\cdot)$ denotes the cosine similarity, and $\tau$ is a temperature hyperparameter. The weighting term $w_{ik}$ is introduced to reduce the penalty for ambiguous pairs:
		\begin{equation}
			w_{ik} = 
			\begin{cases}
				1, & \text{if } c_{ik} \geq t_{\text{pos}} \text{ or } c_{ik} \leq t_{\text{neg}} \\
				\gamma, & \text{if } t_{\text{neg}} < c_{ik} < t_{\text{pos}}
			\end{cases}
		\end{equation}
		where $\gamma \in [0, 1)$ is a hyperparameter.  This design enables the model to learn a robust mapping from channel characteristics to a latent space that preserves the expected structural similarity among beamformers.
	\end{itemize} 
	Then, the final loss for the representation model combines the VAE objective with this new loss:
	\begin{equation}
		\mathcal{L}_{\text{rep}}(\phi,\psi,\zeta) = \mathcal{L}_{\text{vae}} + \beta_1 \mathcal{L}_{\text{ds}} + \beta_2 \mathcal{L}_{\text{ccs}}
	\end{equation}
	where $\beta_1$ and $\beta_2$ are the balancing hyperparameters.
	
	\subsubsection{Critic Network Update}
    To ensure stable policy learning, the centralized critic architecture employs twin critics, $Q^{(1)}$ and $Q^{(2)}$, following the design principle of the TD3 algorithm~\cite{Fujimoto2018}. This twin critic design helps to reduce Q-value overestimation, thereby allowing for a more accurate value estimate of a joint state-action pair. The critic's function is to evaluate the expected total team reward. Therefore, the output of each critic network $Q^{(i)}$ is the sum of the individualized Q-values, $Q^{(i)}_{t}$, from the attention mechanism for agent $t$:
	\begin{equation}
		Q^{(i)}(\mathcal{O}, Z; \theta_Q^{(i)}) = \sum_{t=1}^{K} Q^{(i)}_{t}(\mathcal{O}, Z; \theta_Q^{(i)}), ~\forall i\in\{1,2\},
	\end{equation}
	where $\mathcal{O}=\{\mathcal{O}_1, \dots, \mathcal{O}_K\}$ and $Z=\{z_1, \dots z_K\}$. 
	In the CTDE paradigm, agents act sequentially and the centralized critic evaluates their joint outcome; thus the target value function is the team reward $\mathcal{R}$, defined by the total transmission time $T$, without any bootstrapped next-state value. The loss function for the critic's parameters $\theta_Q^{(i)}$ is thus the MSE between the predicted total Q-value and the actual reward obtained by \eqref{eq:reward}: 
	\begin{equation}
		\mathcal{L}_{\text{cri}}(\theta_Q^{(i)}) = \mathbb{E} \left[ \left( \sum_{t=1}^{K} Q^{(i)}_{t}(\mathcal{O}, Z; \theta_{Q}^{(i)}) - \mathcal{R} \right)^2 \right], ~\forall i\in\{1,2\}.
	\end{equation}
	
	\subsubsection{Actor Network Update}
	Each decentralized actor $\pi_t$ with parameters $\theta_t$ is updated to produce actions that maximize the team's expected return, as estimated by the critic. Following the TD3 algorithm, we use the output of the first critic, $Q^{(1)}$, to guide the policy update. The total actor loss is the negative of the critic's estimated total Q-value:
	\begin{equation}
		\mathcal{L}_{\text{act}}(\theta_{\pi}) = - \mathbb{E}_{\mathcal{O} \sim \mathcal{B}} \left[ Q^{(1)}(\mathcal{O}, \pi_1(\mathcal{O}_1), \dots, \pi_K(\mathcal{O}_K)) \right]
	\end{equation}
	When this combined loss is backpropagated, PyTorch's autograd engine ensures that the parameters $\theta_t$ of each actor $\pi_t$ are updated only with respect to the gradient flowing from their corresponding individualized Q-value, $Q_t^{(1)}$. This allows for a single, efficient backward pass while ensuring each actor receives a precise and personalized learning signal reflecting its own contribution.
	
	%%%%%%%%%%%%%%%%%%%%%%%%%%%%%%%%%%%%%%%%%%%%%%%%%%%%%%%%%%%%%%%%%%%%%%%%%%%%%%

	\subsubsection{Training Procedure}
	%%%%%%%%%%%%%%%%%%%%%%%%%%%%%%%%%%%%%%%%%%%%%%%%%%%%%%%%%%%%%%%%%%%%%%%%%%%%%%
	\begin{algorithm}[t]
		\caption{Training Procedure for the Proposed MARL}% with Hybrid Action Representation and Attention}
		\label{alg:proposed}
		\begin{algorithmic}[1]
			\STATE Initialize actors $\{\pi_{\theta_t}\}$, critics $\{Q^{(k)}\}$, representation models $(E_{\zeta}, q_{\phi}, p_{\psi})$, target networks, and a replay buffer $\mathcal{B}$.
			
			%\STATE \COMMENT{--- Stage 1: Representation Warm-up ---}
			\STATE Populate $\mathcal{B}$ using a random policy and pre-train representation models $(\zeta, \phi, \psi)$ by minimizing $\mathcal{L}_{\text{rep}}$.
			
			%\STATE \COMMENT{--- Stage 2: MARL Training ---}
			\FOR{episode = $1, \dots, N_\text{ep}$}
			\STATE Observe initial observations $\mathcal{O} \leftarrow \{\mathcal{O}_t\}_{t=1}^K$.
			\STATE Select latent action $Z \leftarrow \{\pi_t(\mathcal{O}_t) + \epsilon\}_{t=1}^K$ with exploration noise $\epsilon$ and decode to original action $\{\mathcal{A}_t\}_{t=1}^K$.
			\STATE Execute $\{\mathcal{A}_t\}_{t=1}^K$, observe team reward $\mathcal{R}$, and store $(\mathcal{O}, Z, \mathcal{R})$ in $\mathcal{B}$.
			
			\STATE Sample a minibatch from $\mathcal{B}$.
			\STATE Update critic parameters $\{\theta_{Q}^{(i)}\}_{i=1}^2$ by minimizing $\mathcal{L}_{\text{cri}}$.
			
			\IF {episode \% $N_\text{act}$ == 0}
			\STATE Update actor parameters $\{\theta_{\pi_t}\}_{t=1}^K$ by ascending the policy gradient from $\mathcal{L}_{\text{act}}$.
			\STATE Update all target networks: $\theta' \leftarrow \rho \theta + (1-\rho)\theta'$.
			\ENDIF
			
			\IF {episode \% $N_\text{rep}$ == 0}
			\STATE Update representation models $(\zeta, \phi, \psi)$ by minimizing $\mathcal{L}_{\text{rep}}$.
			\ENDIF
			\ENDFOR
		\end{algorithmic}
	\end{algorithm}
	%%%%%%%%%%%%%%%%%%%%%%%%%%%%%%%%%%%%%%%%%%%%%%%%%%%%%%%%%%%%%%%%%%%%%%%%%%%%%%
	Training has two stages: representation warm-up and MARL training. In the warm-up stage, the hybrid action representation models (embedding table $E_{\zeta}$ and the conditional VAE ($q_\phi, p_\psi$)) are pre-trained. Specifically, the cVAE is trained in a supervised manner, using reference beamformers from a low-complexity solver (e.g., SDR) as a target for accurate reconstruction. This minimizes the $\mathcal{L}_{\text{rep}}$ loss (VAE reconstruction and contrastive loss) using a replay buffer populated by a random policy.  Following this, the main MARL training stage commences, operating within the CTDE paradigm. Decentralized actors generate latent actions from local observations. These actions are decoded into the hybrid action space to interact with the environment. Resulting transitions are stored in the replay buffer. Centralized critics and decentralized actors are then updated via the TD3 algorithm. Details are in Algorithm \ref{alg:proposed}.
	
	\subsection{Network Configurations and Complexity Analysis}
	
	The proposed MARL framework consists of three main components: a representation model for handling the hybrid action space, decentralized actor networks, and a centralized critic network. To analyze the complexity, we first define the dimensions of the key variables. Let $l_{\mathcal{O}}$ be the dimension of an agent's local observation $\mathcal{O}_t$, which includes the request status, local CSI, and cached information matrix. The latent action $z_t$ is composed of a discrete part $z_t^d$ and a continuous part $z_t^c$, with dimensions $l_{z,d}$ and $l_{z,c}$, respectively. The full latent action dimension is $l_z = l_{z,d} + l_{z,c}$. The continuous action, i.e., the beamforming vector $v_{\mathcal{I}_t}$, has a maximum dimension of $l_v = 2 \cdot N_t^2$ (due to the real and imaginary parts). The detailed network structures are outlined in Table \ref{tab:netw_config}.
	\begin{table*}[t]
		\centering
		\caption{Network Configurations of the Proposed MARL Framework.}
		\label{tab:netw_config}
		\begin{tabular}{c|c|c|c}
			%\toprule
			\hline
			\textbf{Component} & \textbf{Sub-Component} & \textbf{Layer} & \textbf{Structure} \\
			%\midrule \midrule
			\hline
			\hline
			% --- Representation Model Section ---
			\multirow{13}{*}{\textbf{Representation Model}} & Discrete Action Embedding Table ($E_\zeta$) & Parameterized Table & $(\mathbb{R}^{D}, \mathbb{R}^{l_{z,d}})$ \\

			\cline{2-4} % & \multirow{6}{*}
			& \multirow{5}{*}{cVAE Encoder ($q_\phi$)} & Fully Connected (Input) & $(\mathbb{R}^{l_v}, 1024)$ \\
			& & Fully Connected (Condition) & $(\mathbb{R}^{l_\mathcal{O} + l_{z,d}}, 1024)$ \\
			& & Element-wise Product \& Activation & ReLU \\
			& & Fully Connected \& Activation & $(1024, 1024)$ / ReLU \\
			%& & Activation & ReLU \\
			& & Fully Connected (Mean \& Log-Std) & $(1024, \mathbb{R}^{l_{z,c} + l_{z,c}})$ \\
			
			\cline{2-4} % & \multirow{6}{*}
			& \multirow{5}{*}{cVAE Decoder ($p_\psi$)} & Fully Connected (Latent) & $(\mathbb{R}^{l_{z,c}}, 1024)$ \\
			& & Fully Connected (Condition) & $(\mathbb{R}^{l_\mathcal{O} + l_{z,d}}, 1024)$ \\
			& & Element-wise Product \& Activation & ReLU \\
			& & Fully Connected \& Activation & $(1024, 1024)$ / ReLU \\
			%& & Activation & ReLU \\
			& & Fully Connected (Reconstruction) & $(1024, \mathbb{R}^{l_v})$ \\
			%\midrule\midrule
			\hline
			\hline
			% --- Actor Network Section --- \multicolumn{2}{c|}{\multirow{6}{*}
			\multicolumn{2}{c|}{\multirow{3}{*}{\textbf{Decentralized Actor ($\pi_t$)}}} & Fully Connected \& Activation  & $(l_\mathcal{O}, 1024)$ / ReLU \\
			%\multicolumn{2}{c|}{} & Activation & ReLU \\
			\multicolumn{2}{c|}{} & Fully Connected \& Activation & $(1024, 1024)$ / ReLU \\
			%\multicolumn{2}{c|}{} & Activation & ReLU \\
			\multicolumn{2}{c|}{} & Fully Connected & $(1024, \mathbb{R}^{l_{z,d}+l_{z,c}})$  \\
			%\multicolumn{2}{c|}{} & Activation & None \\
			\hline
			\hline
			%\midrule\midrule
			% --- Centralized Critic Section ---\multicolumn{2}{c|}{\multirow{6}{*}
			\multicolumn{2}{c|}{\multirow{4}{*}{\textbf{Centralized Critic ($Q$)}}} & Fully Connected (Embedding $f_\epsilon$) & $(l_\mathcal{O} + l_z, 128)$ \\
			\multicolumn{2}{c|}{} & Attention Mechanism ($M_\mathbf{q}, M_\mathbf{k}, M_\mathbf{w}$) & Query, Key, Value Projections from $\mathbb{R}^{128}$ \\
			\multicolumn{2}{c|}{} & Fully Connected \& Activation & $(128+128, 128)$ / ReLU \\
			%\multicolumn{2}{c|}{} & Activation & ReLU \\
			\multicolumn{2}{c|}{} & Fully Connected (Output $f_Q$) & $(128, 1)$  \\
			%\multicolumn{2}{c|}{} & Activation & None \\
			%\bottomrule
			\hline
		\end{tabular}
	\end{table*}
	
	\subsubsection{Inference Complexity}
	During decentralized execution, each of the $K$ agents runs its actor network and the cVAE decoder to determine its hybrid action. 
	Assuming the number of layers is fixed and comparable in size, we approximate the per-pass complexity as proportional to the product of the input and output dimensions \cite{Sipper1993}. 
	For a single agent, the forward pass through the actor network mapping the local observation $\mathcal{O}_t$ to the latent action $z_t$ has a complexity of approximately $O(l_{\mathcal{O}} \cdot l_z)$. 
	The discrete action is then obtained via a nearest-neighbor search in the embedding table, with complexity $O(D \cdot l_{z,d})$. 
	Finally, the cVAE decoder generates the continuous beamforming vector, with complexity $O((l_z + l_{\mathcal{O}}) \cdot l_v)$. 
	The total inference complexity for one time step is:
	\begin{equation}
		O\left(K \cdot \left(l_\mathcal{O} \cdot l_z + D \cdot l_{z,d} + (l_z + l_\mathcal{O}) \cdot l_v\right)\right).
	\end{equation}
	
	\subsubsection{Training Complexity}
	In centralized training, the per-step cost includes the cVAE encoder/decoder, all $K$ actor networks, and the centralized critic with an attention mechanism over all agents. 
	Assuming the number of layers is fixed and comparable in size, and applying the same complexity approximation as in \cite{Sipper1993}, the representation model (cVAE encoder and decoder) has a complexity of roughly $O((l_v + l_{\mathcal{O}} + l_{z,d}) \cdot l_{z,c} + (l_z + l_{\mathcal{O}}) \cdot l_v)$. 
	The centralized critic, which processes inputs from all $K$ agents, uses an attention mechanism with complexity $O(K^2 \cdot l_{\varepsilon})$ for pairwise agent relationships. 
	Each actor network is updated from its individualized Q-value gradients. 
	Summing these components, the overall training complexity per step is:
	\begin{equation}
		O\left( (l_v+l_{\mathcal{O}} + l_{z,d})l_{z,c} + (l_z+l_{\mathcal{O}})l_v + K^2 l_{\varepsilon} + K(l_{\mathcal{O}}+l_z)l_{\varepsilon} \right),
	\end{equation}
	which shows that the decentralized execution complexity scales linearly in $K$, while centralized training scales quadratically due to the attention mechanism.
	
	Ultimately, unlike conventional optimization methods that require exhaustive EIC searches and iterative beamformer updates, the learned MARL policy enables rapid inference. This approach relies solely on local observations for decentralized execution, thereby eliminating burdensome centralized computations. This efficiency makes the framework particularly effective for the continuously arriving tasks and dynamic channel conditions inherent in modern distributed computing systems\footnote{Coded shuffling, used beyond MapReduce, now aids multi-task learning by sharing intermediate gradients, showing coding frameworks' relevance to modern distributed machine learning \cite{Hu2023}.}.

	%%%%%%%%%%%%%%%%%%%%%%%%%%%%%%%%%%%%%%%%%%%%%%%%%%%%%%%%%%%%%%%%%%%
	% % % % % % % % % % % % % % % % % % % % % % % % % % % % % % % % % %
	%%%%%%%%%%%%%%%%%%%%%%%%%%%%%%%%%%%%%%%%%%%%%%%%%%%%%%%%%%%%%%%%%%%
	\section{Numerical Results} 
	\label{sec:NUM}
	In this section, we evaluate the performance of our proposed joint index coding and beamforming design by assessing the total transmission time. We conduct simulations by randomly generating cached information scenarios for all users and averaging the results over 30 independent trials. Unless stated otherwise, the system parameters are set as follows: the number of users is $K=5$, each equipped with $N_t=4$ antennas, the transmit power budget for each user $P=1$, and the file size for each transmission is $B=10^5$ bits.
	For the proposed MARL framework, additional hyperparameters are specified. We use a batch size of 32, with learning rates for the actor-critic networks and the cVAE set to $1 \times 10^{-5}$ and $1 \times 10^{-3}$, respectively. The dimensions for the discrete and continuous latent actions, $l_{z,d}$ and $l_{z,c}$, are both 32. The dimension of the continuous action vector $l_v$, representing the real and imaginary parts of the beamformers, is $2 \cdot N_t^2$. For exploration, we add noise $\epsilon$ from a normal distribution with a mean of 0, where the standard deviation linearly anneals from 1.0 to 0.05 over the first 1,000 episodes. The soft target network update parameter $\rho$ is $5 \times 10^{-3}$. The models are trained for $N_\text{ep} = 2,000$ episodes, with actor updates every $N_\text{act} = 2$ episodes and representation model updates every $N_\text{rep} = 50$ episodes with 10 updates each. The loss-balancing hyperparameters $\beta_1$ and $\beta_2$ are set to 2 and 1, and the temperature $\tau$ is 0.1.  The parameters $t_\text{pos}$ and $t_\text{neg}$, and $\gamma$ are set to 0.88, 0.6, and 0.1, respectively. To stabilize the training of the cVAE, we employ KL annealing. This technique gradually increases the weight of the KL divergence term in the VAE loss, which prevents the model from ignoring the input data (known as posterior collapse) and allows it to learn more meaningful representations.

	%%%%%%%%%%%%%%%%%%%%%%%%%%%%%%%%%%%%%%%%%%%%%%%%%%%%%%%%%%%%%%%%%%%%%%%%%%%%%%
	\begin{figure}[t]
		\centering
		\begin{subfigure}[b]{0.68\columnwidth}
			\includegraphics[width=\linewidth]{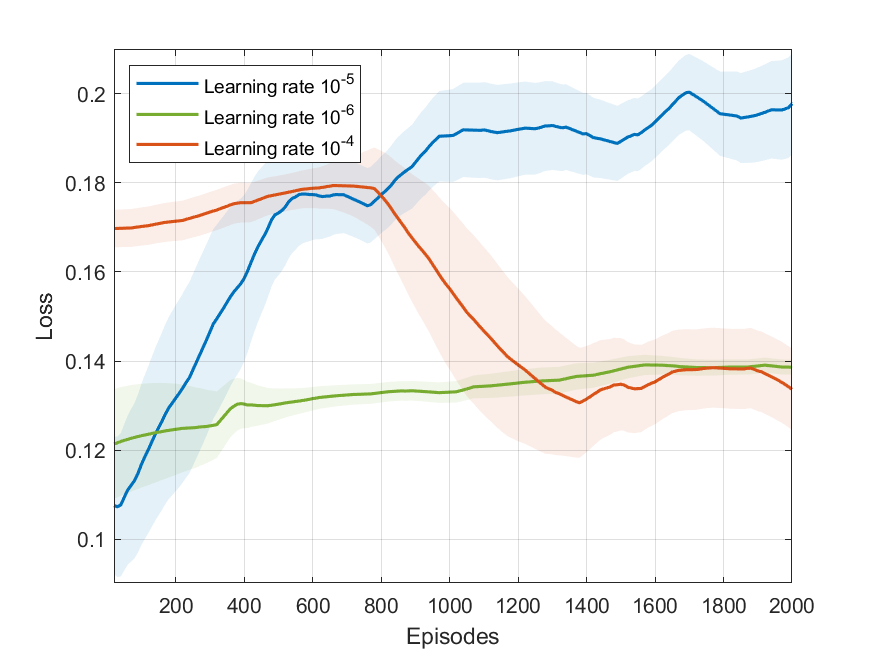}
			\caption{}%Convergence w.r.t. learning rates.}
			\label{fig:lr}
		\end{subfigure}
		%\hfill 
		\begin{subfigure}[b]{0.68\columnwidth}
			\includegraphics[width=\linewidth]{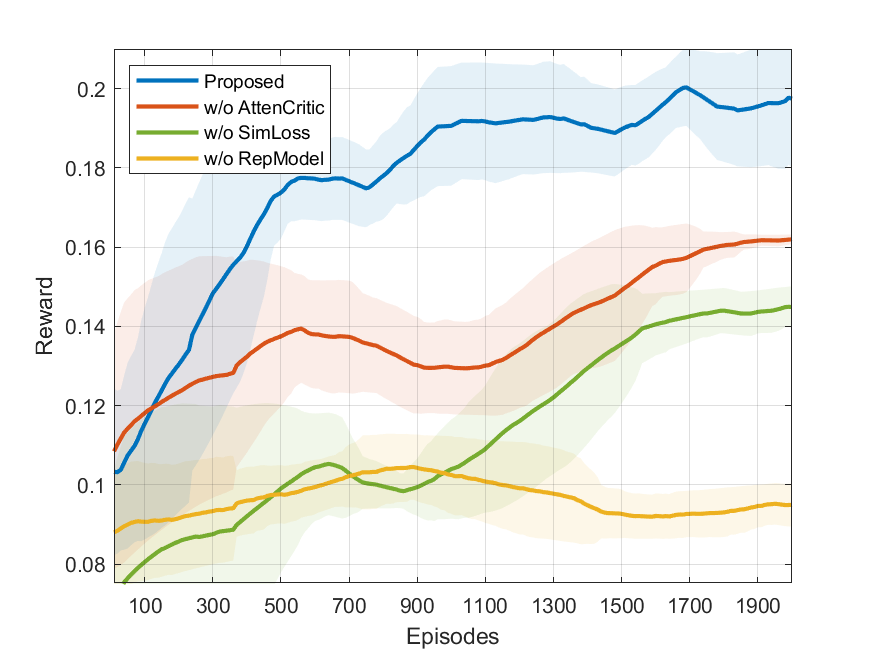}
			\caption{}%Ablation study.}
			\label{fig:ablation}
		\end{subfigure}
		
		\caption{Convergence plots: (a) Different learning rates and (b) Ablation study results.}
		\label{fig:conv} 
	\end{figure}
	%%%%%%%%%%%%%%%%%%%%%%%%%%%%%%%%%%%%%%%%%%%%%%%%%%%%%%%%%%%%%%%%%%%%%%%%%%%%%%
	Fig. \ref{fig:conv} illustrates the convergence behavior of the proposed MARL framework. Fig. \ref{fig:conv}(a) shows the training reward over 2,000 episodes for different learning rates. A higher learning rate of $10^{-4}$ leads to faster initial learning but suffers from high variance and converges to a suboptimal policy. Conversely, a smaller learning rate of $10^{-6}$ demonstrates stable and monotonic improvement but converges much more slowly. A learning rate of $10^{-5}$ provides a balanced trade-off, achieving near-optimal performance with rapid and stable convergence. Fig. \ref{fig:conv}(b) presents an ablation study, highlighting the importance of each component in our MARL design. Removing the attention-based critic (\textit{w/o AttenCritic}) or the specialized representation model (\textit{w/o RepModel}) results in a significant performance degradation, underscoring their critical roles in effective credit assignment and handling the hybrid action space, respectively. Furthermore, removing the similarity loss (\textit{w/o SimLoss}), which structures the latent space, also degrades the final converged performance, suggesting that placing semantically similar actions closer together contributes to learning a more refined policy.
	
	%%%%%%%%%%%%%%%%%%%%%%%%%%%%%%%%%%%%%%%%%%%%%%%%%%%%%%%%%%%%%%%%%%%%%%%%%%%%%%
	\begin{figure}[t]
		\centering
		{\includegraphics[width=1\columnwidth]{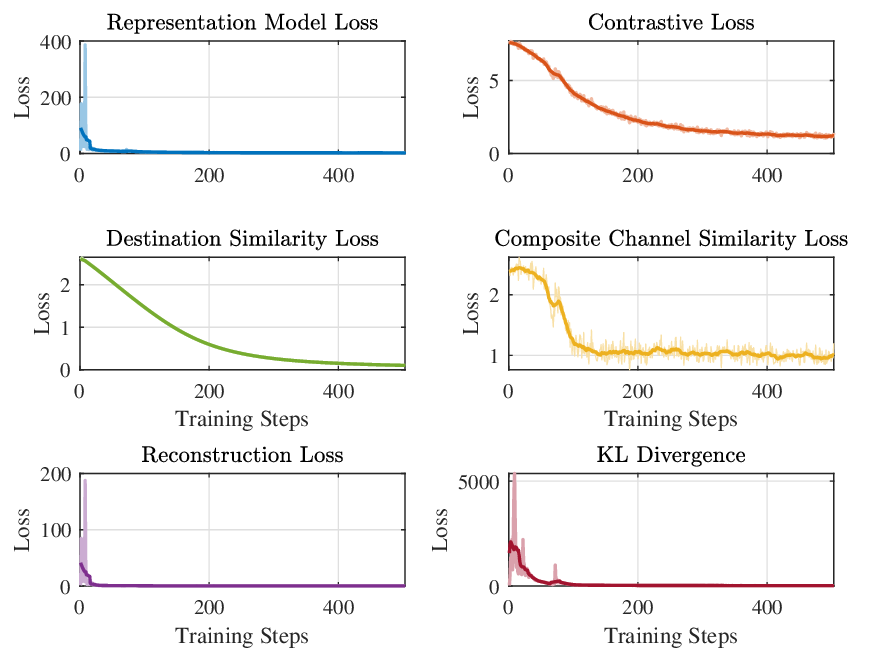}} 
		\caption{\!\!\! The training losses of the representation model.}
		\label{fig:vaeloss}
	\end{figure}
	%%%%%%%%%%%%%%%%%%%%%%%%%%%%%%%%%%%%%%%%%%%%%%%%%%%%%%%%%%%%%%%%%%%%%%%%%%%%%%
	Fig. \ref{fig:vaeloss} details the training losses for the hybrid action representation model during the initial warm-up phase. The plots show that all loss components, which include the overall representaion model loss $\mathcal{L}_\text{rep}$, the overall contrastive loss (weighted sum of $\mathcal{L}_\text{ds}$ and $\mathcal{L}_\text{ccs}$), and their constituent parts like reconstruction loss and KL divergence, steadily decrease over training steps. This confirms the representation model (a cVAE and embedding table) is effectively trained, having learned a structured and smooth latent space to accurately map discrete EIC choices and continuous beamforming vectors, which is a crucial prerequisite for the subsequent MARL policy training.
	
	%%%%%%%%%%%%%%%%%%%%%%%%%%%%%%%%%%%%%%%%%%%%%%%%%%%%%%%%%%%%%%%%%%%%%%%%%%%%%%
	\begin{figure*}[t]
		\centering
		\begin{subfigure}[b]{0.315\textwidth} %0.38
			\includegraphics[width=\linewidth]{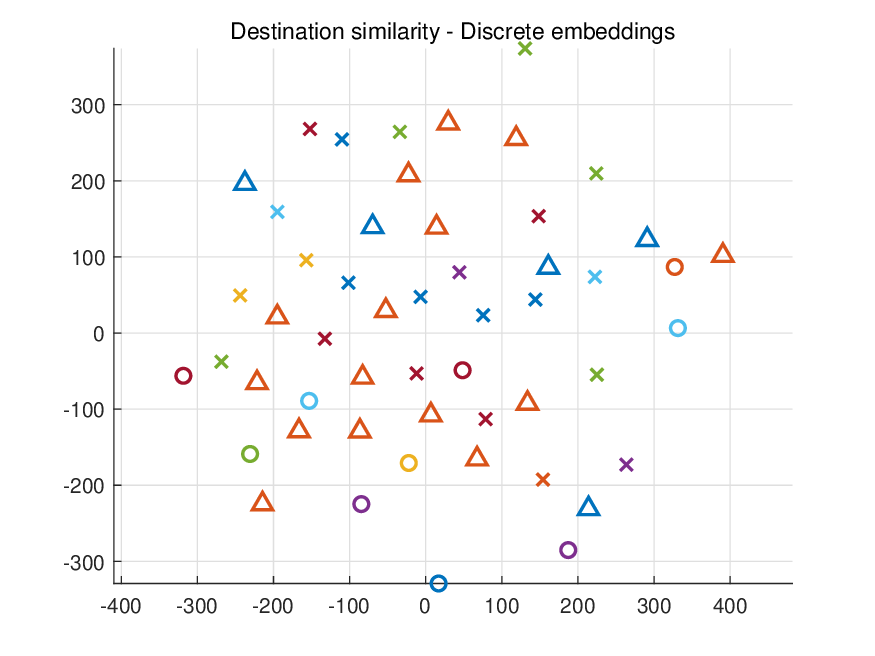}
			\caption{}
			\label{fig:ds_before}
		\end{subfigure}
		%\hfill
		\begin{subfigure}[b]{0.315\textwidth}
			\includegraphics[width=\linewidth]{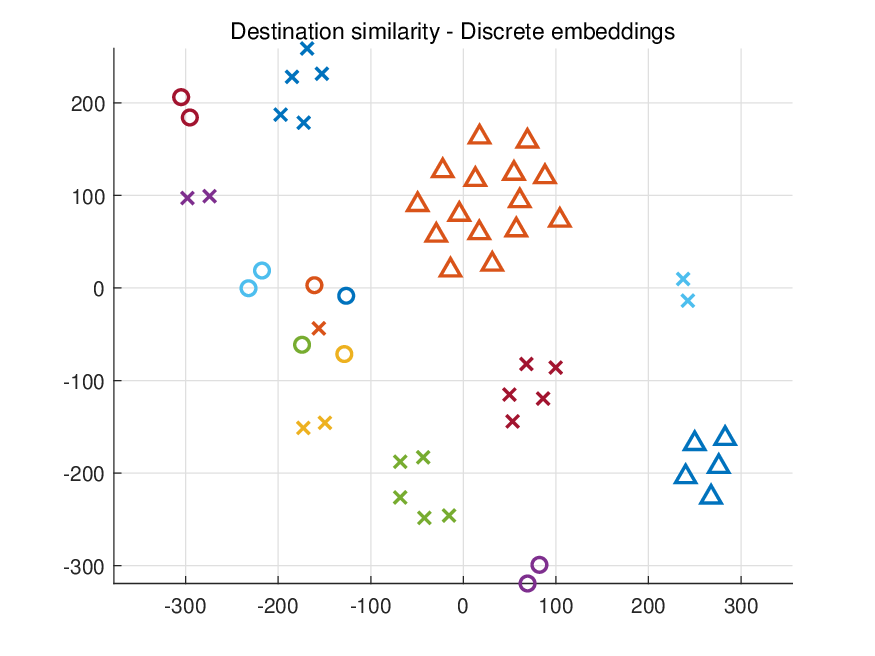}
			\caption{}
			\label{fig:ds_after}
		\end{subfigure}
		%\hfill
        \\
		\begin{subfigure}[b]{0.315\textwidth}
			\includegraphics[width=\linewidth]{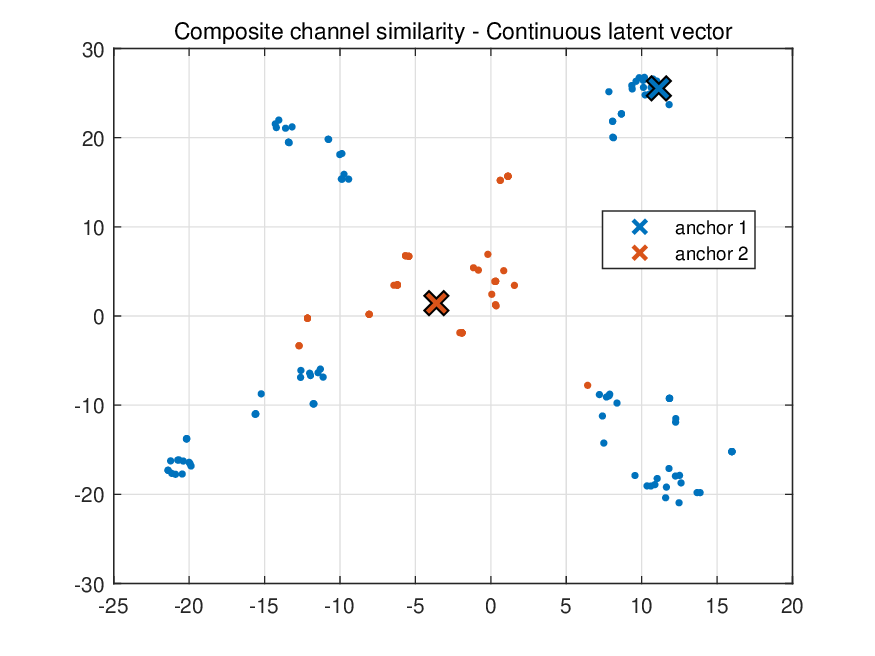}
			\caption{}
			\label{fig:ccs_before}
		\end{subfigure}
		%\hfill
		\begin{subfigure}[b]{0.315\textwidth}
			\includegraphics[width=\linewidth]{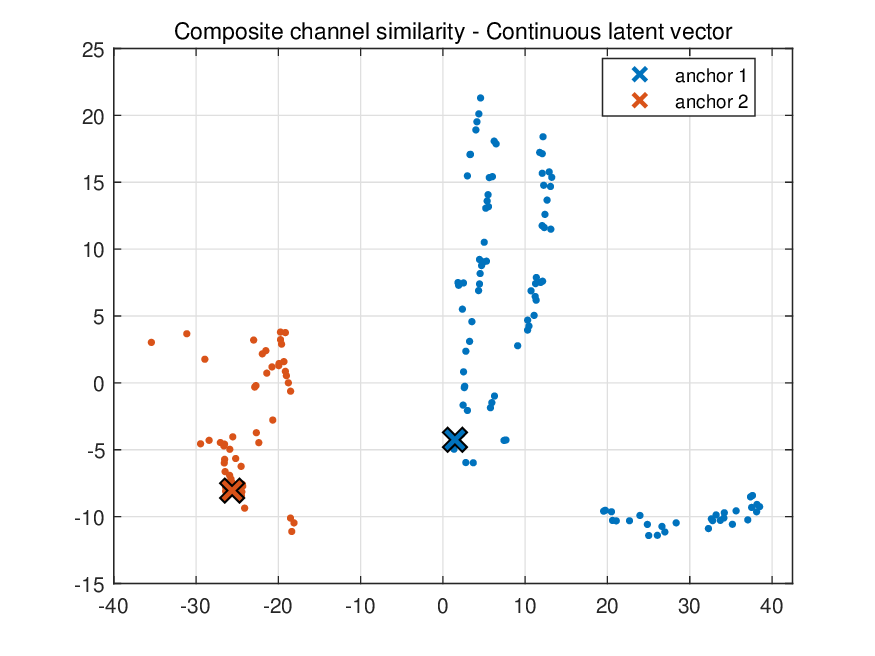}
			\caption{}
			\label{fig:ccs_after}
		\end{subfigure}
		
		\caption{The t-SNE plots before and after training: (a) discrete embeddings (before), (b) discrete embeddings (after), (c) continuous latent vector (before), and (d) continuous latent vector (after).}
		\label{fig:tsne}
		\vspace{-10pt}
	\end{figure*}
	
	%%%%%%%%%%%%%%%%%%%%%%%%%%%%%%%%%%%%%%%%%%%%%%%%%%%%%%%%%%%%%%%%%%%%%%%%%%%%%%
    Fig. \ref{fig:tsne} provides a visual representation of the learned latent space using t-distributed Stochastic Neighbor Embedding (t-SNE) plots, both before and after training the representation model, where t-SNE is a nonlinear dimensionality reduction technique that projects high-dimensional data into a low-dimensional space for visualization while preserving local structure.  Identically colored and shaped markers indicate actions that either target the same set of destination users ((a) and (b)) or exhibit high composite channel similarity to one of two selected anchor samples ((c) and (d)). Initially, the discrete action embeddings are scattered randomly without any discernible structure, as seen in Fig. \ref{fig:tsne}(a). After training, however, the model learns to group actions based on semantic similarity. Fig.~\ref{fig:tsne}(b) shows that discrete actions targeting the same set of destination users are clustered together, guided by the destination similarity loss. A similar transformation occurs in the continuous latent space. Fig. \ref{fig:tsne}(c) visualizes the continuous latent vectors from 200 random batches, which also appear randomly distributed before training. Correspondingly, Fig. \ref{fig:tsne}(d) shows these vectors now clearly separated into distinct clusters around the two anchors. Points are colored by their composite channel similarity to anchors, confirming our loss organizes the latent space by grouping vectors that require structurally similar beamformers. This vital structure allows the policy network to generalize smoothly across similar actions, leading to more efficient and stable learning.
    
    As seen in Fig.~\ref{fig:tsne}(a), the discrete action embeddings are initially scattered without any discernible structure. After training, however, the model learns to group actions based on semantic similarity, and Fig.~\ref{fig:tsne}(b) shows these embeddings tightly clustered by their common destination users, guided by the destination similarity loss. A similar transformation occurs in the continuous latent space. Fig.~\ref{fig:tsne}(c) visualizes the continuous latent vectors from 200 random batches, which also appear randomly distributed before training. Correspondingly, Fig.~\ref{fig:tsne}(d) shows these same vectors now clearly separated into distinct clusters around two anchor samples. 
	%%%%%%%%%%%%%%%%%%%%%%%%%%%%%%%%%%%%%%%%%%%%%%%%%%%%%%%%%%%%%%%%%%%%%%%%%%%%%%
	\begin{figure}[t]
		\centering
		{\includegraphics[width=0.68\columnwidth]{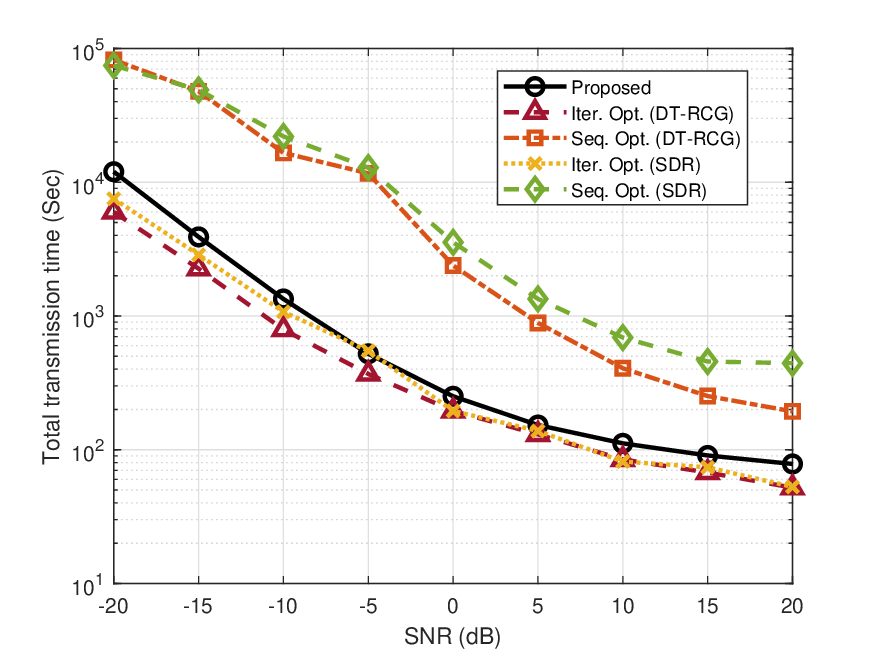}} % sim8_1
		\caption{\!\!\! The total transmission time with respect to the SNR.} 
		\label{fig:snr}
	\end{figure}
	%%%%%%%%%%%%%%%%%%%%%%%%%%%%%%%%%%%%%%%%%%%%%%%%%%%%%%%%%%%%%%%%%%%%%%%%%%%%%%
	
	Fig. \ref{fig:snr} compares the total transmission time of the proposed MARL-based approach against several conventional optimization-based benchmarks across a range of transmit SNRs. The benchmarks include both a joint iterative optimization (\textit{Iter. Opt.}) as in Section \ref{sec:CON} and a separate sequential optimization (\textit{Seq. Opt.}), which first finds the minimum-length EIC and subsequently designs the beamformer\footnote{To our knowledge, we are the first to address the WEIC problem in a MIMO setting. We validated our approach's effectiveness by comparing it against the joint iterative optimization, which performs close to optimally.}. To solve the complex multi-group multicast beamforming subproblem within these benchmarks, we implement two distinct solvers: the \textit{DT-RCG} method introduced in the Section \ref{sec:CON} and a widely-used Semidefinite Relaxation (\textit{SDR}) method \cite{Huang2010}. The results show that for both the iterative and sequential settings, the DT-RCG consistently achieves slightly better performance than the SDR approach, highlighting the efficacy and slight advantage of DT-RCG. 
	
	The proposed MARL method consistently outperforms the sequential optimization schemes (both \textit{Seq. Opt. (DT-RCG)} and \textit{Seq. Opt. (SDR)}). This highlights the significant gains achieved by jointly optimizing the network-layer index coding and physical-layer beamforming. Furthermore, our approach achieves a total transmission time that is nearly identical to the iterative, centralized optimization baselines (\textit{Iter. Opt.}) that rely on exhaustive search and global CSI for all D2D pairs. This near-optimal performance is achieved with substantially reduced complexity and without the need for global CSI at execution time. While a minor performance gap exists at very low SNRs, the MARL approach's ability to make rapid, decentralized decisions based only on local observations makes it a far more practical solution for dynamic wireless environments where centralized information gathering is infeasible.  
	
	%%%%%%%%%%%%%%%%%%%%%%%%%%%%%%%%%%%%%%%%%%%%%%%%%%%%%%%%%%%%%%%%%%%%%%%%%%%%%%
	\begin{figure}[t]
		\centering
		\begin{subfigure}[b]{0.68\columnwidth}
			\includegraphics[width=\linewidth]{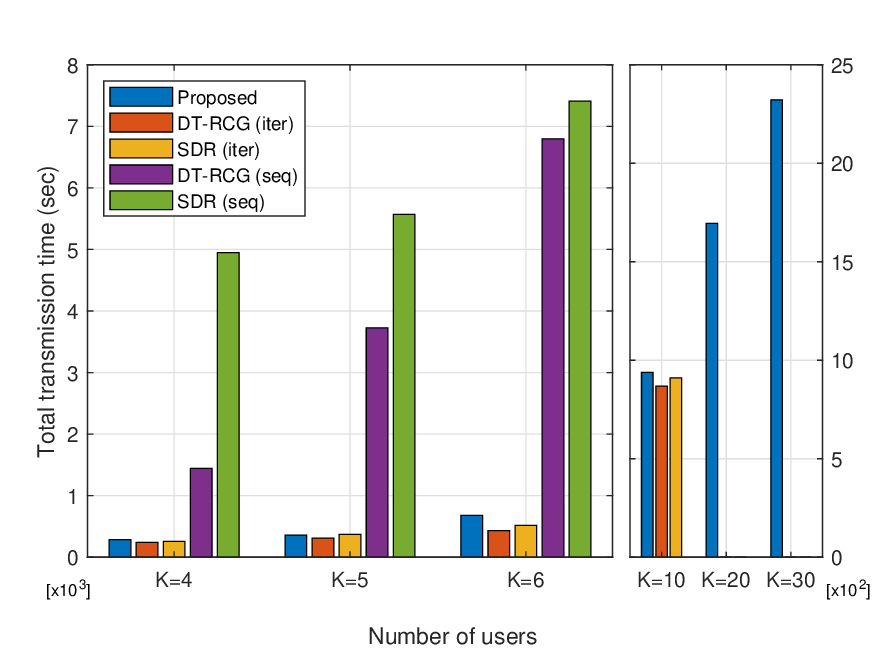}
			\caption{}
			\label{fig:k}
		\end{subfigure}
		%\hfill 
		\begin{subfigure}[b]{0.68\columnwidth}
			\includegraphics[width=\linewidth]{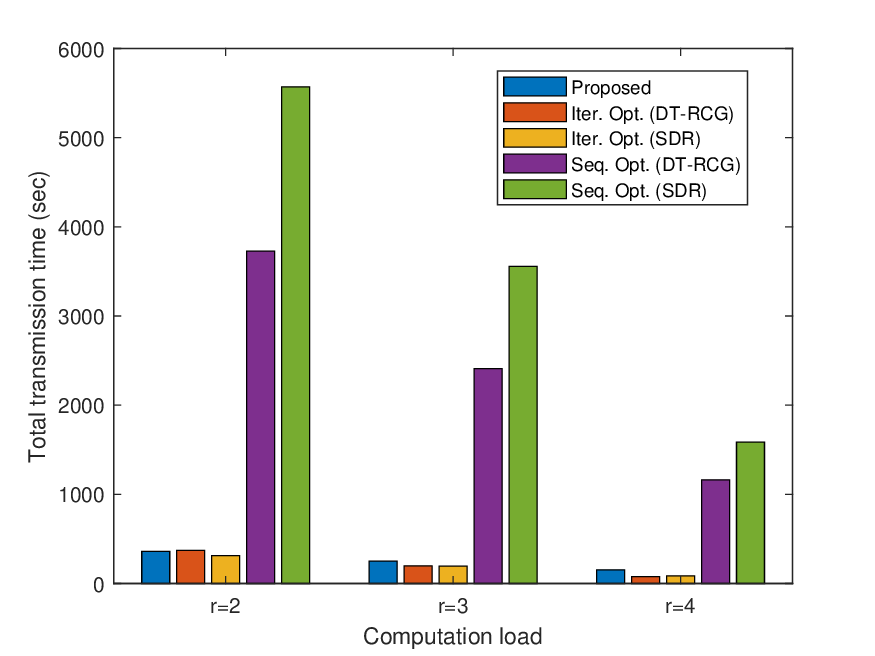}
			\caption{}%Ablation study.}
			\label{fig:r}
		\end{subfigure}
		
		\caption{The total transmission time with respect to (a) the number of users $K$ and (b) the computation load $r$.}
		\label{fig:kr} 
	\end{figure}
	%%%%%%%%%%%%%%%%%%%%%%%%%%%%%%%%%%%%%%%%%%%%%%%%%%%%%%%%%%%%%%%%%%%%%%%%%%%%%%

	Fig.~\ref{fig:kr} further evaluates the proposed design with respect to both the number of users ($K$) and the computation load ($r$).
The computation load $r \triangleq \|\mathcal{M}\|_1 / K$ quantifies the average amount of cached data per user, serving as a proxy for the available local computation results before shuffling.%
\footnote{In MapReduce-like distributed computing, higher computation loads reduce shuffling and increase coding opportunities \cite{Song2022}. Similarly, in EIC, larger cache sizes are positively correlated with more coding opportunities.}
	    
    The left subfigure of Fig.~\ref{fig:kr}(a) illustrates the performance of the proposed MARL framework, where the model is trained independently for each scenario with $K=4,~5$, and $6$. As shown, the total transmission time increases moderately with larger $K$, primarily because more users lead to more data to be delivered. Nevertheless, both the proposed MARL and the \textit{Iter.~Opt.} methods effectively balance coding and spatial multiplexing gains, keeping the growth in transmission time relative to \textit{Seq.~Opt.} small and maintaining strong performance across all tested $K$ values.
        
      To further address the scalability challenges of MARL in large-scale systems, user clustering can be employed. This strategy aligns naturally with our motivating distributed computing scenario, where a master node can strategically manage data placement and redundant computations across worker nodes. Such control enables the formation of worker groups with high side-information overlap, resulting in clusters rich in coding opportunities. Applying a pre-trained policy to these smaller, well-defined clusters provides a computationally efficient approach to scaling our framework.
      
      The right subfigure of Fig.~\ref{fig:kr}(a) validates this strategy in larger systems with $K=10$, $K=20$, and $K=30$, where a pre-trained model for $K=5$ is applied to two, four, and six distinct five-user clusters, respectively. Notably, the clustering strategy can be adapted to environmental factors, for instance, grouping users by spatial proximity can enhance channel quality, highlighting the flexibility of our approach. For $K=10$, the results show that our method maintains only a marginal performance gap compared with the \textit{Iter.~Opt.} baselines, which rely on exhaustive search. However, for $K=20$ and $K=30$, \textit{Iter.~Opt.} solutions could not be obtained due to the intractable computational complexity of exhaustive search. This observation is consistent with the known NP-hardness of the EIC problem, where finding an optimal solution becomes practically impossible as the network scale increases \cite{Porter2021, Son2019, Son2021}. In contrast, the proposed MARL framework finds a solution and, by leveraging user clustering, maintains a roughly linear increase in transmission time, proving its scalability and effectiveness in large-scale networks.  The \textit{Seq.~Opt.} results are omitted, as their poor performance required a different plot scale. Overall, these results confirm the practicality of using user clustering to extend pre-trained models to larger systems. However, the specific clustering method significantly influences outcomes, making it crucial to select one that aligns with environmental characteristics to maintain performance gains.
      	
    Meanwhile, Fig.~\ref{fig:kr}(b) shows that as $r$ increases, the total transmission time decreases across all schemes. This is because a larger cache size in EIC provides more coding opportunities, enabling multiple users to be served simultaneously. Across all tested $r$ values, the proposed MARL method consistently matches the performance of the computationally intensive iterative optimization baselines, demonstrating its ability to fully exploit the additional coding opportunities afforded by larger cache sizes.

    Taken together, these results show that the proposed decentralized policy achieves near-optimal performance relative to centralized optimization while avoiding the heavy runtime of high-complexity iterative methods. Its robustness to varying system parameters, such as the number of workers and the computation load, demonstrates that the learned policy generalizes to diverse practical scenarios. The proposed approach is a strong foundation for designing efficient and scalable distributed computing systems, where resources can be flexibly adapted without significant shuffling penalties, enabling faster job completion and more efficient resource usage.
	
	%%%%%%%%%%%%%%%%%%%%%%%%%%%%%%%%%%%%%%%%%%%%%%%%%%%%%%%%%%%%%%%%%%%
	\section{Conclusion} \label{sec:CONC}	
    This work investigated the joint design of embedded index coding and beamforming for MIMO-based wireless D2D communications in distributed computing, to minimize the total transmission time. We proposed a two-step optimization method and extended it with a distributed MARL framework that effectively operates on local observations under hybrid discrete-continuous action constraints. The proposed MARL approach alleviates the prohibitive complexity of exhaustive search while adapting to dynamic wireless environments. Numerical results demonstrate that the proposed framework achieves significant performance gains with low computational complexity. These findings indicate that combining embedded index coding with spatial multiplexing, optimized through learning-based techniques, offers a scalable and efficient solution to the WEIC problem in distributed computing systems.
    
	%%%%%%%%%%%%%%%%%%%%%%%%%%%%%%%%%%%%%%%%%%%%%%%%%%%%%%%%%%%%%%%%%%%
	% % % % % % % % % % % % % % % % % % % % % % % % % % % % % % % % % %
	%%%%%%%%%%%%%%%%%%%%%%%%%%%%%%%%%%%%%%%%%%%%%%%%%%%%%%%%%%%%%%%%%%%

\end{document}